\documentclass[aps, pra, twocolumn,10pt,longbibliography]{revtex4-1}

\usepackage{amsmath, amssymb, amsfonts, amsthm, bm, mathrsfs, bbm, dsfont}
\usepackage{graphicx}
\usepackage{epstopdf}
\usepackage{color}
\usepackage[usenames,dvipsnames]{xcolor}
\usepackage[citecolor=blue, colorlinks=true, urlcolor=blue, linkcolor=blue]{hyperref}

\newcommand{\ket}[1]{|#1\rangle}
\newcommand{\bra}[1]{\langle#1|}

\DeclareMathOperator{\tr}{\mbox{tr}}

\DeclareMathOperator{\cerf}{cerf}

\newcommand{\ve}[1]{{\bf #1}}

\newcommand{\eq}[1]{Eq.~(\ref{#1})}

\newcommand{\fig}[1]{Fig.\thinspace{}\ref{#1}}

\newcommand{\fc}[1]{({#1})}
\newcommand{\figc}[2]{Fig.\thinspace{}\ref{#1}\thinspace{}\fc{#2}}

\newcommand{\identity}{\hat{\mathds{1}}}

\usepackage{times}

\begin{document}

\title{Entanglement production and information scrambling in a noisy spin system}

\author{Michael Knap}

\affiliation{Department of Physics and Institute for Advanced Study, Technical University of Munich, 85748 Garching, Germany}

\begin{abstract}
	
We study theoretically entanglement and operator growth in a spin system coupled to an environment, which is modeled with classical dephasing noise. Using exact numerical simulations we show that the entanglement growth and its fluctuations are described by the Kardar-Parisi-Zhang equation. Moreover, we find that the wavefront in the out-of-time ordered correlator (OTOC), which is a measure for the operator growth, propagates linearly with the butterfly velocity and broadens diffusively with a diffusion constant that is larger than the one of spin transport. The obtained entanglement velocity is smaller than the butterfly velocity for finite noise strength, yet both of them are strongly suppressed by the noise. We calculate perturbatively how the effective time scales depend on the noise strength, both for uncorrelated Markovian and for correlated non-Markovian noise. 

\end{abstract}

\date{\today}

\pacs{
}

\maketitle

\section{Introduction}

One of the major challenges in quantum statistical physics is to understand the fundamental principles of the thermalization dynamics in isolated quantum many-body systems~\cite{deutsch_quantum_1991, srednicki_chaos_1994, rigol_thermalization_2008}. An essential part of it is the irreversible growth of quantum information, that is quantified by the von Neumann entanglement entropy, as a complex quantum many-body system evolves in time~\cite{calabrese_evolution_2005, kim_ballistic_2013, mezei_entanglement_2017-1, nahum_quantum_2017, zhou_emergent_2018}. Due to the unitarity of the time evolution, quantum information in the initial state is never truly lost. However, it gets scrambled in non-local correlations, which can be accessed by the out-of-time ordered correlators (OTOCs)~\cite{larkin_1969}. OTOCs have been explored in field theories~\cite{kitaev_2014, shenker_black_2014, hosur_chaos_2016, polchinski_spectrum_2016, maldacena_bound_2016, roberts_diagnosing_2015, stanford_many-body_2016, maldacena_bound_2016, patel_quantum_2017, werman_quantum_2017, gu_local_2017, jian_universal_2018}, in the semi-classical limit~\cite{larkin_1969, aleiner_microscopic_2016, kurchan_quantum_2016, rozenbaum_lyapunov_2017, rammensee_many-body_2018, khemani_velocity-dependent_2018}, and in lattice systems~\cite{bohrdt_scrambling_2017, chen_out--time-order_2017, luitz_information_2017, kukuljan_weak_2017, jonay_coarse-grained_2018, pappalardi_scrambling_2018}. These quantities are not only of theoretical interest. Due to the unprecedented level of control in synthetic quantum matter, entanglement entropies~\cite{islam_measuring_2015, kaufman_quantum_2016} and information scrambling~\cite{garttner_measuring_2017, li_measuring_2017, landsman_verified_2018} has been measured experimentally in complex many-body systems. Recently, there has been significant progress in analytically understanding these phenomena in random circuit models, which consist of structureless Haar random gates that are applied stochastically in space and time~\cite{nahum_quantum_2017, nahum_operator_2018, von_keyserlingk_operator_2018, khemani_operator_2018, rakovszky_diffusive_2018, chan_solution_2017, gopalakrishnan_operator_2018,  hamma_quantum_2012, brown_decoupling_2015}. However, it remains a challenge to understand the generic aspects of operator and entanglement growth in  conventional many-body models.

\begin{figure}[h!]
	\includegraphics[width=.48\textwidth]{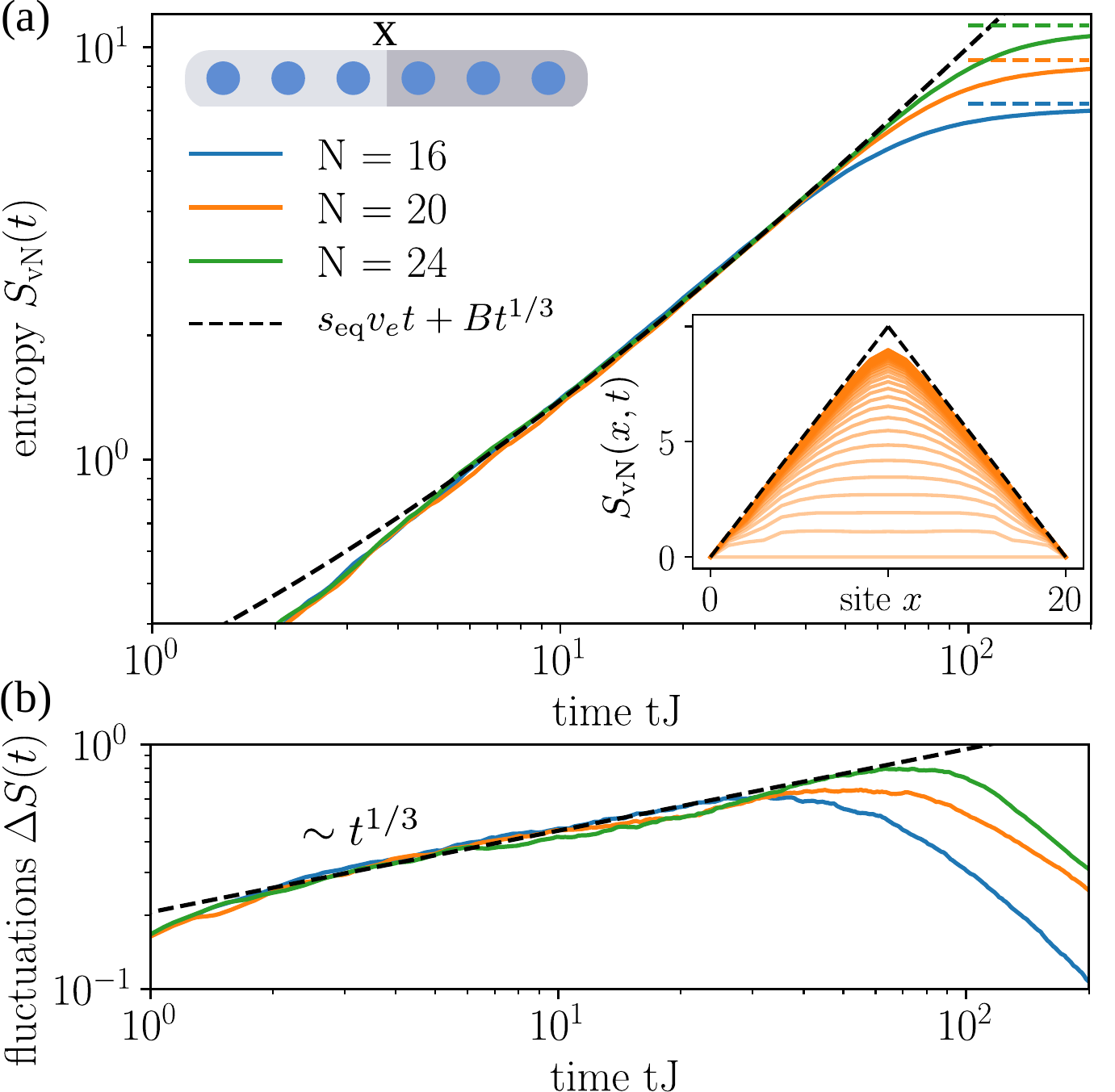}
	\caption{\textbf{Entanglement production in a noisy spin chain.} We compute the von Neumann entanglement entropy for a Heisenberg chain with $N$ spins starting with initial unentangled product states. In the due course of the dynamics the spins are subjected to dephasing noise of amplitude $\Lambda = 3\sqrt{J}$. The entanglement cut is chosen to separate the system into two equal halves. \fc{a} The von Neumann entanglement entropy obeys a KPZ scaling with a strong sublinear power-law correction $S_\text{vN} = s_\text{eq} v_e t + B t^{1/3}$, dashed line. Inset: The entanglement growth  for cuts at different positions $x$ (later times are indicated by more intense colors). 
	\fc{b} As a consequence of the KPZ scaling, the entanglement fluctuations also scale as a powerlaw with exponent ${1/3}$. At late times the entanglement starts to saturate and hence the powerlaw scaling in the fluctuations breaks down as well.
	}
	\label{fig:entanglement}
\end{figure}

In the present work, we study the entanglement production and information scrambling in a paradigmatic Heisenberg spin chain subject to dephasing noise. The effect of noise is to act as a bath on the spins which slows down the quantum dynamics. Because of this slow down, we can efficiently calculate the asymptotic behavior of the operator and entanglement growth using exact numerical techniques for  moderately sized systems. We find that the exponents for entanglement production in our system are those of the Kadar-Parisi-Zhang (KPZ) equation, which has been originally introduced for stochastic surface growth~\cite{kardar_dynamic_1986}. This scaling implies that the mean of the entropy grows linearly with time, i.e., we can associate a rate (or 'velocity') with the entanglement production. By contrast, the fluctuations scale with a nontrivial powerlaw exponent, see \fig{fig:entanglement}. 
We furthermore, calculate the OTOC of the quantum spins and find that its wavefront can be captured by a biased random walk distribution, i.e., it propagates ballistically with the butterfly velocity and spreads diffusively in time.
Therefore, our result show that the phenomenology predicted from random unitary circuits in the large local Hilbert space limit in which random gates are applied sequentially at discrete times steps~\cite{nahum_quantum_2017, zhou_emergent_2018, rakovszky_diffusive_2018, nahum_operator_2018}, holds also for noisy spin-1/2 systems evolved continuously in time. In the strong noise limit, we can calculate analytically the effective times scale that governs entanglement and operator growth and find that it is proportional to the noise strength. As a consequence, both the entanglement and the butterfly velocity depend strongly on the noise with the former being strictly smaller than the latter.

As our work was nearing completion two related studies on operator growth in noisy systems appeared~\cite{xu_locality_2018, rowlands_noisy_2018}. Our work differs from Ref.~\onlinecite{xu_locality_2018} in that it also considers coherent dynamics and from both works in that ours does not necessarily rely on the Markovian white noise limit.

\section{A noisy spin system}We consider a Heisenberg spin chain, subjected to time dependent noise, as described by
\begin{equation}
 \hat H_t = - J \sum_j [\vec{\hat{S}}_j \cdot \vec{\hat{S}}_{j+1} + \frac{\delta}{2}(S^+_j S^-_{j+2} + \text{h.c.}) ]+ \sum_j \xi_j(t) \hat S^z_j,
 \label{eq:H}
\end{equation}
where we explicitly denote the time dependence of the Hamiltonian by a subscript $t$. In our model, $J$ is the strength of the Heisenberg coupling of neighboring spins, $\delta$ characterizes the next-to-nearest neighbor flip-flop processes, which we introduce to break the integrability of the Heisenberg model, and $\xi_j(t)$ is white noise with amplitude $\Lambda$ (with units of $(\text{energy})^{1/2}$), i.e., 
\begin{equation}
\langle \xi_i(t) \xi_j(t')\rangle = \Lambda^2 \delta(t-t') \delta_{ij}.
\label{eq:noise}
\end{equation}
However, in general, also correlated, non-Markovian noise may be considered by introducing a finite noise correlation time, see e.g. Refs.~\onlinecite{gopalakrishnan_noise-induced_2017, amir_classical_2009}. We will be interested in how the entanglement production and the spreading of operators changes as a function of the noise amplitude $\Lambda$.

When averaging over all noise trajectories, this model can for white noise be mapped onto a Lindblad master equation with jump operators that are given by $\hat S_j^z$~\cite{breuer_theory_2002}:
\begin{equation}
\frac{\partial \hat \rho}{\partial t} = \mathcal{L} \hat \rho  \equiv -i [\hat H, \hat \rho] + \Lambda^2 \sum_j [\hat S_j^z \hat \rho \hat S_j^z - \frac14 \hat \rho],
\label{eq:L}
\end{equation}
where $\mathcal{L}$ is the Lindblad superoperator, with a coherent part that generates the Heisenberg time evolution  ($\hat H$ is the time-independent part of \eq{eq:H}) and an incoherent part that leads to dephasing. 

In the large noise limit $\Lambda^2/J \gg 1$, we show using perturbation theory that both the entanglement  and the operator growth are governed by an effective timescale 
\begin{equation}
\tau \sim \Lambda^2/J^2.
\label{eq:timescale}
\end{equation} 
From that we obtain that diffusion constants, entanglement velocities, and butterfly velocities are parametrically suppressed with noise. Thus only moderately sized systems are required to observe the universal behavior of entanglement production and operator spreading, which makes it favorable to study this model using exact numerical techniques.  We compute the dynamics of our system, fixing $\delta = 0.5$ for various values of the noise amplitude $\Lambda$ using exact Krylov time evolution of \eq{eq:H} and perform the average over noise configurations explicitly.

\section{Entanglement production}Starting with an unentangled product state $\ket{\psi}$, we evolve our system in time $\ket{\psi(t)} = U_\xi(t)\ket{\psi}$ under the unitary  $U_\xi(t) = \mathcal{T} \exp[-i \int_0^t \hat H_\tau d\tau]$. At each time step, we bipartition our systems at position $x$ and trace out the subsystem right to the cut (c.f. \figc{fig:entanglement}{a} top left inset). This leaves us with a density matrix of the left subsystem, $\hat \rho_x(t)$. From this density matrix we compute the von Neumann entanglement entropy
\begin{equation}
S_\text{vN}(x,t) = -\tr [\hat \rho_x(t) \log \hat \rho_x(t)].
\end{equation}
We choose the basis of the logarithm to be two, so that the maximum entanglement of two spins counts one unit. Time traces of $S_\text{vN}(t)$ averaged over several hundreds of noise configurations are shown in \figc{fig:entanglement}{a} for up to 24 spins. Additional data for other noise strengths and also for R\'enyi entropies are shown in App.~\ref{app:A}.

\begin{figure*}
	\includegraphics[width=\textwidth]{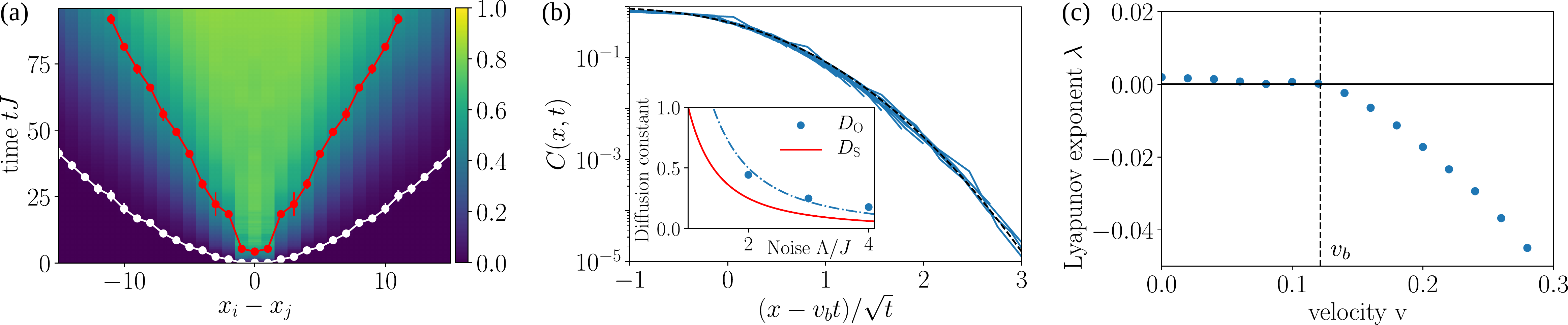}
	\caption{\textbf{Operator scrambling.} We calculate the out-of-time ordered correlator (OTOC) $C(x_i-x_j,t)$ for noise amplitude $\Lambda = 3 \sqrt{J}$. \fc{a} A contour plot of the OTOC is shown in addition to the contour line at which the OTOC takes half (red curve) and one percent (white curve) of its saturation value. Errorbars indicate the standard error of the mean. The slope of the linear red curve defines the inverse butterfly velocity $v_b$. \fc{b} Scaling collapse over five orders of magnitude of the diffusively broadening right wavefront of the OTOC, which moves with the butterfly velocity $+v_b$. The inset shows the spin diffusion constant $D_\text{S}$, red line, and the diffusion constant of the wavefront $D_{O}$. Both diffusion constants scale as $1/\Lambda^2$. \fc{c} The velocity dependent Lyapunov exponent $\lambda(v)$, blue dots, becomes negative at the butterfly velocity $v_b$.
	}
	\label{fig:otoc}
\end{figure*}

The entanglement dynamics has been recently computed for analytically tractable  models based on random unitary circuits with local Hilbert space dimension $q$ (our case of spin-1/2s corresponds to $q=2$)~\cite{nahum_quantum_2017, zhou_emergent_2018}. Using the constraints from subadditivity of the von Neumann entropy, the entanglement production could be mapped in the limit of a large local Hilbert space ($q\gg 1$) to a classical growth model which obeys KPZ scaling
\begin{equation}
 S_\text{vN}(t) = s_\text{eq} v_e t + B t^{1/3},
 \label{eq:KPZ}
\end{equation}
where the first term signals linear growth of the entanglement  and the second term is a sublinear correction with a nontrivial powerlaw determined from the KPZ solution~\cite{prahofer_universal_2000,prahofer_statistical_2000}. We can interpret the leading linear entanglement growth also as velocity $v_e$ by noting that the entanglement between two consecutive sites in space is the equilibrium entropy $s_\text{eq}$ in the steady state. A noisy system approaches infinite temperature at late times, hence $s_\text{eq} = \log(2)$. For times $tJ \gtrsim 5$, our entanglement production data \figc{fig:entanglement}{a} is consistent with this form with a rather large prefactor $B$ of the sublinear term which depends on the noise strength, App.~\ref{app:A}. 
At shorter times $tJ \lesssim 5$ the data does not obey KPZ scaling, resulting from the crossover of the short-time entanglement dynamics, for which interactions are not relevant, to the long-time KPZ scaling established by interactions, App.~\ref{app:B}. The inset of \figc{fig:entanglement}{a} shows the entanglement entropy for different cuts in space at position $x$. At early times the average of the entropy grows uniformly for all cuts, whereas it saturates to the pyramid shaped maximal entanglement (dashed lines) at late times. 

To substantiate the KPZ scaling, we analyze the temporal fluctuations of the entanglement entropy $\Delta S(x, t) = [\langle (S_\text{vN}(t)- \langle S_\text{vN}(x, t) \rangle)^2 \rangle]^{1/2}$; here $\langle \cdot \rangle$ represents both an average over noise trajectories and initial product states. The numerically evaluated entanglement fluctuations scale as a nontrivial powerlaw with exponent $1/3$, and are therefore consistent with the KPZ scaling of the fluctuations. At late times this scaling breaks down because the entanglement saturates to a finite value. We also find the KPZ exponent of $1/2$ for spatial fluctuations of the entanglement entropy (not shown). Our results thus show that temporal noise is sufficient for the entanglement growth to obey KPZ scaling, similarly as in random unitary circuits, even though the latter are locally structureless and evolve at discrete time steps.

\section{Strong noise limit}The strong noise limit admits a perturbative treatment. We separate the Lindbladian $\mathcal{L}$, \eq{eq:L}, into the coherent contribution $\mathcal{{L}}_1 \hat \rho = -i [\hat H, \hat \rho]$ and the incoherent contribution $\mathcal{{L}}_0 \hat \rho = \sum_j [\hat S_j^z \hat \rho \hat  S_j^z - \frac14 \hat \rho]$. In the strong noise limit, we calculate the effects of $\mathcal{L}_1$ perturbatively. The steady states of the unperturbed (dissipative) term $\mathcal{{L}}_0$ are of the form $\ket{\ve{s}}\bra{\ve{s}}$, where $\ve{s}$ is a string of spin states in the basis of $\hat S^z$. All of these states have eigenvalue $\lambda_0=0$. We can now employ second-order perturbation theory to obtain an effective Lindblad operator
\begin{equation}
\mathcal{{L}}_\text{eff} = \mathcal{\hat{P}} \mathcal{{L}}_1 \frac{1}{\lambda_0 - \mathcal{{L}}_0} \mathcal{{L}}_1 \mathcal{\hat{P}}
\label{eq:Leff}
\end{equation}
where $\mathcal{\hat{P}}$ projects onto the subspace of spanned by the eigenstates of $\mathcal{{L}}_0$~\cite{cai_algebraic_2013,lesanovsky_kinetic_2013}. In the strong noise limit, we use $\mathcal{{L}}_\text{eff}$ to predict the dissipative dynamics.

We will first study time-ordered correlation functions. In our model the total spin $\sum_i \hat S_i^z$ is conserved. Hence, long wavelength excitations of such conserved operators follow hydrodynamics, which manifests as a diffusive mode at long wavelengths $\omega(k) = -i D_{S} k^2 + \ldots$, where the dots represent corrections that are of higher order in $k$~\cite{chaikin_principles_2000}. Writing down the equation of motion for $\hat S^z_k$ (by replacing $\hat \rho$ with $\hat S^z_k$ and $\hat H$ with $-\hat H$ in \eq{eq:L}) and expanding around small momenta, one obtains the spin diffusion constant~\cite{bauer_stochastic_2017, han_locality_2018} 
\begin{equation}
  D_S = (1+4\delta^2) \frac{J^2}{2 \Lambda^2}.
  \label{eq:DS}
\end{equation}
Hence, the spin diffusion constant is reduced with increasing noise strength~\cite{znidaric_dephasing-induced_2010, han_locality_2018}. In the strong noise limit, this relation also holds for arbitrary anisotropies in the $S^z_j S^z_{j+1}$ term of the Heisenberg model. 

In our system transport is diffusive, however, operator and entanglement growth is ballistic because of the interactions~\cite{bohrdt_scrambling_2017, kim_ballistic_2013}. Using this perturbative approach, we can compute the timescale that limits the rate of the entanglement production and the spreading of the operators (see also Ref.~\onlinecite{rowlands_noisy_2018}). To this end, we consider a Lindblad equation for the operator $\hat \rho \otimes \hat \rho$ on an extended product space. From $\hat \rho \otimes \hat \rho$ we obtain the purity of a subsystem by summing over the proper indices, $\langle{\rho_x^2(t)}\rangle = \langle \sum_{ij} [\rho_x(t)]_{ij} \otimes [\rho_x(t)]_{ji} \rangle$, where the first and second density matrix come from the left and right product space of $\hat \rho \otimes \hat \rho$ and as before $\hat \rho_x$ is the reduced density matrix with spins located at positions right to $x$ being traced out. The entanglement of the purity $S_p = -\log (\langle{\rho_x^2(t)}\rangle)$ differs from the second  R\'enyi entropy $S_2 = -\langle{\log[\tr \rho_x^2(t)]}\rangle$ in the orders of the average. However, we find that both quantities behave similarly, see App.~\ref{app:A}, which is why we assume the timescale $\tau$ obtained from the dynamics of $\hat \rho \otimes \hat \rho$ governs generally the entropy growth. In a related way, the OTOC can be computed with the time evolved operator copied on the two product spaces~\cite{kitaev_2014}. 

The Lindblad equation for $\hat \rho \otimes \hat \rho$ is~\cite{breuer_theory_2002, rowlands_noisy_2018}
\begin{eqnarray}
\frac{\partial (\hat \rho\otimes \hat \rho)}{\partial t} = -i [\mathcal{\hat H}, \hat \rho\otimes \hat \rho] + {\Lambda^2} \sum_j \Big(\mathcal{\hat S}_j^z \hat \rho\otimes \hat \rho \mathcal{\hat S}_j^z \nonumber\\-  \frac12 [(\mathcal{\hat S}_j^z)^2 \hat \rho\otimes \hat \rho - \hat \rho\otimes \hat \rho (\mathcal{\hat S}_j^z)^2] \Big), 
\label{eq:LL}
\end{eqnarray}
where both $\mathcal{\hat H} = \hat H \otimes \identity + \identity \otimes \hat H$ and $\mathcal{\hat S}_j^z = \hat S_j^z \otimes \identity + \identity \otimes \hat S_j^z$ act on the extended space. By performing a perturbative analysis~\cite{cai_algebraic_2013,rowlands_noisy_2018} (see App.~\ref{app:C}), one obtains an effective Lindblad superoperator $\mathcal{{L}}_\text{eff} = -\frac{1}{\Lambda^2} \mathcal{\hat P} {\mathcal{ L}}_1^2$. As a consequence, the effective timescale for transport, operator growth, and entanglement dynamics scale with the noise  $\Lambda^2$; see \eq{eq:timescale}.

\section{Operator scrambling}The scrambling of information can be characterized by the out-of-time ordered correlator (OTOC)~\cite{larkin_1969, kitaev_2014, shenker_black_2014, hosur_chaos_2016}
\begin{equation}
C(x_i,t) = -2 \langle [ \hat S_i^z(t), \hat S_0^z(0) ]^2 \rangle .
\end{equation}
We have introduced the factor of two such that the OTOC grows to the maximal value of one. The OTOC, shown for noise amplitude $\Lambda = 3\sqrt{J}$ in \figc{fig:otoc}{a}, spreads linearly in time and exhibits a pronounced wave-front broadening. From the linear spreading at half of its saturation value, we obtain the butterfly velocity $v_b$ (red line) playing the role of a Lieb Robinson velocity~\cite{lieb_finite_1972}. We demonstrate that the wavefront broadens diffusively by rescaling $x$ as $\frac{x-v_b t}{\sqrt{t}}$, see \fc{b} where we find a scaling collapse over five orders of magnitude. We extract the diffusion constant $D_{O}$ of the wavefront broadening, by fitting the collapsed data to the complementary error function, $ \cerf[(x-v_bt)/\sqrt{4 D_{O}t}]/2$, which is the inverse cumulative distribution function of the Gaussian that governs the biased random walk (dashed black line). The diffusion constant $D_O$ is shown in the inset of \fc{b}, blue symbols, along with spin diffusion constant $D_S$ from \eq{eq:DS}, red line. The operator diffusion is about a factor two larger than spin diffusion constant and also follows a scaling of $1/\Lambda^2$, dashed-dotted blue line, determined by the effective time scale $\tau$, \eq{eq:timescale}. 

From the complementary error function form of the OTOC, we calculate the contour lines at lower saturation values $c$:  $x = v_b t + \cerf^{-1}(2c) \sqrt {4 D t}$. This equation fits well to the numerically obtained contour points, indicated by the white symbols in panel \figc{fig:otoc}{a}. From that it becomes apparent that there is no separate light cone velocity in our model (in contrast to random unitary dynamics where the light cone velocity is defined by the rate at which gates are applied).

Analogously to the analysis of classical chaos~\cite{deissler_one-dimensional_1984, kaneko_lyapunov_1986}, we introduce a velocity dependent Lyapunov exponent $\lambda(v)$ by analyzing the behavior of the OTOC on constant velocity lines (i.e., rays from the origin in \figc{fig:otoc}{a})~\cite{khemani_velocity-dependent_2018, lieb_finite_1972}. In classical systems, the OTOC grows exponentially within the light cone, yet due to the finite local Hilbert space in our quantum system, the OTOC saturates to a finite value, which is why the Lyapunov exponent must be strictly zero in that regime, see \figc{fig:otoc}{c}. At the butterfly velocity $v_b$ (dashed line), $\lambda(v)$ departs from zero and becomes negative indicating an exponential suppression of the OTOC. 

\begin{figure}
	\includegraphics[width=.44\textwidth]{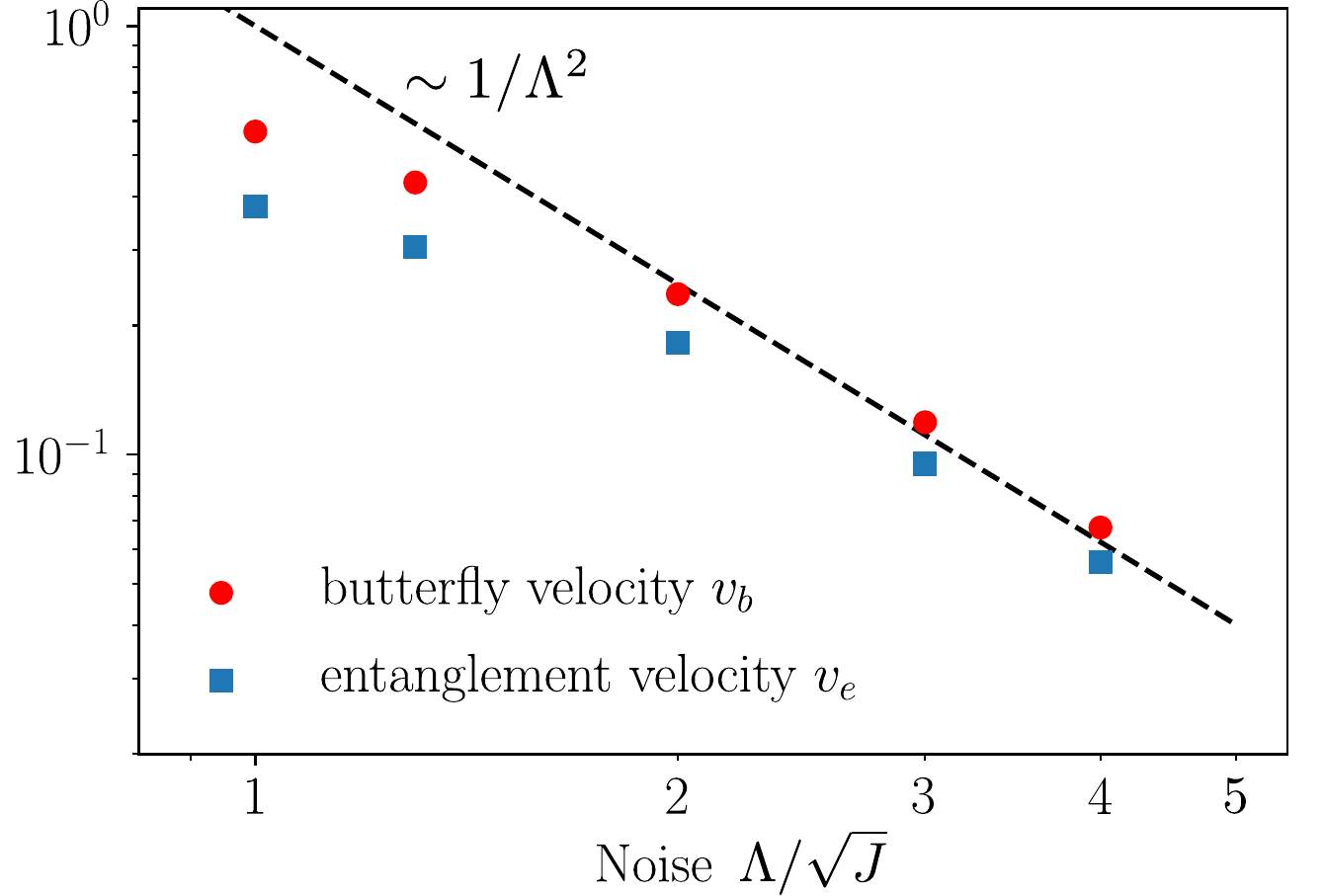}
	\caption{\textbf{Comparison of the entanglement velocity and the butterfly velocity.} The entanglement velocity $v_e$ (blue squares) is strictly  smaller than the butterfly velocity $v_b$ (red dots) for all values of the noise $\Lambda$, even though they approach each other with increasing noise. In the strong noise limit, the velocities are governed by the inverse time scale $\tau^{-1} \sim J^2/\Lambda^2$, which is shown as a black dashed line.
	}
	\label{fig:vel}
\end{figure}
\begin{figure*}
	\includegraphics[width=.98\textwidth]{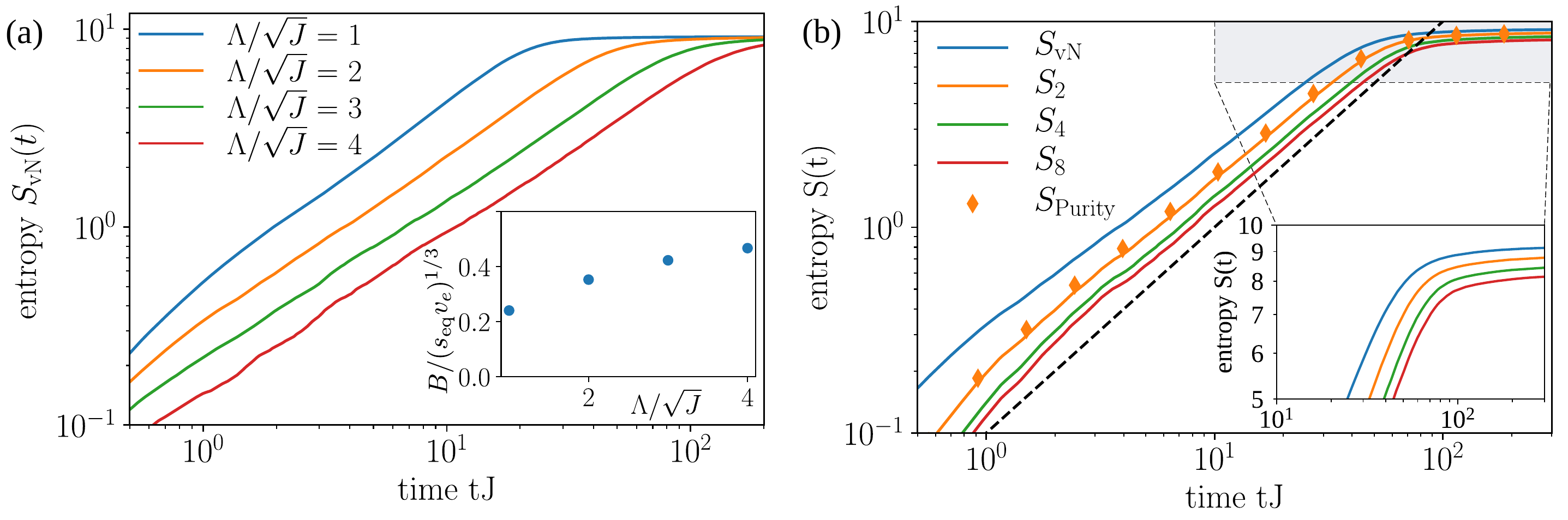}
	\caption{\textbf{ von Neumann, R{\'e}nyi, and purity entanglement entropies.} \fc{a} Entanglement growth for different values of the noise amplitude $\Lambda$ in a spin chain of length $N=20$. Inset: The ratio between the subleading contribution to the entanglement growth and the entanglement growth rate increases with noise. \fc{b} The  R\'enyi entropies $S_n$ are shown for noise amplitude $\Lambda=2\sqrt{J}$  along with the von Neumann entropy and the purity entropy. The purity entropy, diamonds, is very close the second R\'enyi entropy. The subleading corrections $B$ to the entanglement entropies decreases with the index $n$ of the  R\'enyi entropy. Inset: The saturation value of the entropy decreases with increases R\'enyi index $n$.
	}
	\label{fig:renyi}
\end{figure*}

\section{Entanglement vs.\ butterfly velocity}Combining our analysis of the entanglement production and the operator scrambling we compare the two emergent velocities; the entanglement velocity $v_e$ and the butterfly velocity $v_b$. We find from our numerical results that $v_e < v_b$ for all values of the noise $\Lambda$, see \fig{fig:vel}. This inequality results from the diffusive wave-front broadening of the OTOC~\cite{von_keyserlingk_operator_2018}. Since the operator diffusion constant $D_O$ goes to zero with increasing noise, we expect the two velocities to approach each other, as observed in our data. In the strong noise limit, the effective time scale is $\tau \sim \Lambda^2/J^2$. Since the velocities are inversely proportional to time, they scale as $1/\Lambda^2$ for large $\Lambda$, see \fig{fig:vel} where the scaling with $1/\Lambda^2$ is indicated as a black dashed line.

\section{Non-Markovian noise}So far, we have discussed the Markovian white noise limit, in which transport, entanglement, and operator growth are captured by the effective timescale $\tau \sim \Lambda^2/ J^2$. We now argue how this timescale gets modified in the non-Markovian limit in which noise has a finite correlation time. For simplicity we focus on the Ornstein-Uhlenbeck process, $\langle \xi_i(t) \xi_j(0) \rangle = \lambda^2 \exp[-|t|/\sigma] \delta_{ij}$, where $\sigma$ is the noise correlation time and $\lambda$ the noise strength which has units of energy. But other forms of noise can be considered as well. The primary mechanism for spreading of charge and information is the incoherent hopping of spins between neighboring sites. Using a Fermi's Golden Rule-type argument, the effective incoherent hopping rate is~\cite{amir_classical_2009, gopalakrishnan_noise-induced_2017},
$\tau^{-1} = 2J^2 |C_\phi(t)|^2$, 
where $C_\phi(t)  =\int_0^\infty e^{-2 \int_0^{t'} (t-t') \langle \xi(t') \xi(0) \rangle dt'}$. In the limit of fast noise $\lambda \sigma \ll 1$, $C_\phi(t) \sim 1/(2\lambda^2 \sigma)$ and hence 
\begin{equation}
\tau^{-1}_\text{fast} \sim \frac{J^2}{\lambda^2 \sigma} \sim \frac{2J^2}{\int_{-\infty}^\infty \langle \xi_i(t) \xi_j(0) \rangle}.
\end{equation}
Evaluating the last expression for white noise, \eq{eq:noise}, we obtain our typical noise timescale \eq{eq:timescale}. This approach generalizes to slow non-Markovian noise $\lambda \sigma \ll 1$, where $C_\phi(t) \sim C_\phi(0) \sim \exp[-\lambda^2t/2]$. From that we obtain
\begin{equation}
\tau^{-1}_\text{slow} \sim \frac{J^2}{\lambda}. 
\end{equation}
Therefore, we find that entanglement and operator growth are scaling differently in the fast and in the slow noise limit. In the latter case the typical time scale is independent of the precise value of the noise correlation time.

\section{Conclusions and Outlook.}In this work, we have studied information scrambling and operator spreading in a noisy spin system, where the noise models the environment. While it might be expected that noise is detrimental for the entanglement and operator growth, classical noise still retains the unitarity of quantum evolution. Therefore, the main consequence of noise is to increase the effective time scale in the system, which scales with the noise strength. Diffusion constants, butterfly velocities, and  entanglement velocities are thus strongly suppressed with noise. For future work, it will be interesting to numerically explore our conjecture about entanglement and operator growth timescales in the non-Markovian noise limit. Moreover, exciting open questions are what the effect of integrability is on information scrambling and operator growth, and how non-hermitian Lindblad jump operators destroy quantum coherence and what the timescales for such processes are and.  
 
 \appendix
 
 \section{Additional data on the entanglement entropies \label{app:A}}

 Additional data on the growth of the von Neumann entanglement entropy is shown in \figc{fig:renyi}{a} for different values of the noise amplitude $\Lambda$ and for systems of 20 spins. With increasing noise the entanglement growth gets suppressed. The subleading term $B$ gains weight relatively to the entanglement velocity, inset.
 
 R\'enyi entropies generalize the von Neumann entropy and are defined as
 \begin{equation}
 S_{n}(x,t) = \langle \frac{1}{1-n} \log [\tr (\rho_x^n(t))] \rangle
 \end{equation}
 In the limit $n \to 1$, the  R\'enyi entropy reduces to the von Neumann entropy. In \fig{fig:renyi} we show the von Neumann entropy along with  R\'enyi entropies with index $n=2,4,8$ for noise amplitude $\Lambda=2 \sqrt{J}$. The subleading contribution to the entanglement production becomes weaker for increasing  R\'enyi index. In addition the saturation value of the  R\'enyi entropies decreases with  R\'enyi index, see inset in \fig{fig:renyi}, which is consistent with recent results on random unitary circuit models~\cite{zhou_emergent_2018}, in which a generalized Page formula for the saturation value has been derived $S_n^\text{Page} = \frac{N}{2} \log 2 - \log C_n / (n-1)$, where $C_n$ is the n-th Catalan number. 
 We have also computed the Purity entropy,  $S_\text{Purity} = \log [\langle \tr(\rho_x^2(t)) \rangle]$, where the trace and the average over noise is exchanged compared to the second R\'enyi entropy. Both of these entropies are very close to each other. 
 
 \section{Short time dynamics of the entanglement entropy \label{app:B}}
 
\begin{figure}
 	\includegraphics[width=.48\textwidth]{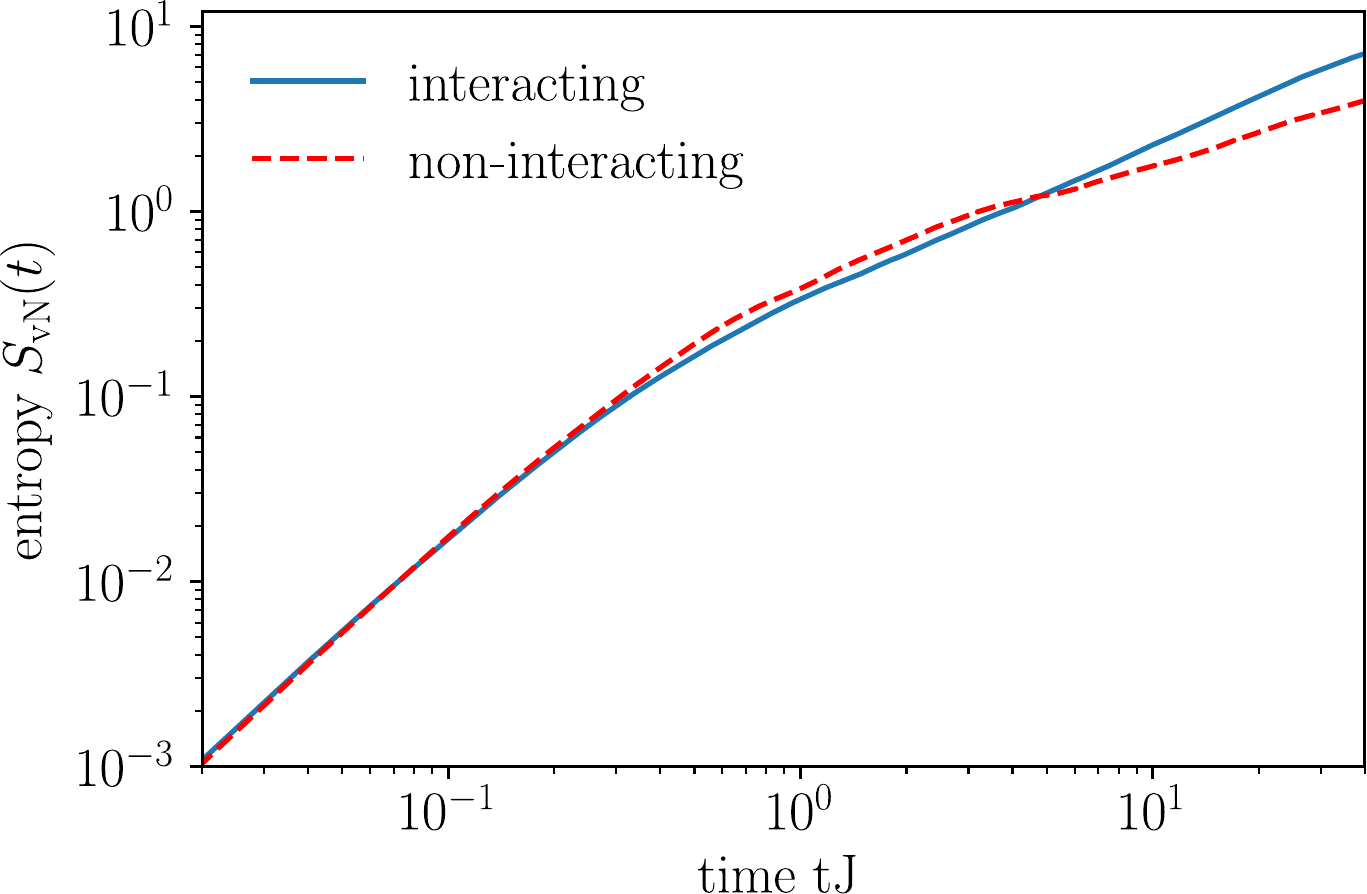}
 	\caption{\textbf{ Comparison of the entanglement dynamics in interacting and non-interacting systems.} We compare the entanglement dynamics of the interacting spin chain studied in the main body of this work with the one of free fermions for noise amplitude $\Lambda = 2 \sqrt{J}$. At short time $tJ \lesssim 1$ both results agree well, indicating that interactions are not relevant for the initial build up of the entanglement. At late time, the free fermion entanglement crosses over to a $\sqrt t$ behavior, whereas, interactions in the spin chain lead to the KPZ entanglement growth. The short time free fermion evolution, and the late time KPZ evolution of the entanglement are separated by a crossover regime.
 	}
 	\label{fig:free}
 \end{figure}

 At short times, interactions are not relevant for the entanglement dynamics. We can therefore understand the short time dynamics by studying the corresponding model of noisy free fermions. We compute the entanglement growth of the free fermion system using the techniques developed in Ref.~\cite{peschel_calculation_2003}. The free fermion entanglement dynamics is compared to the exact Heisenberg dynamics in \fig{fig:free} for noise amplitude $\Lambda = 2\sqrt{J}$. For times $tJ \lesssim 1/2$ both results agree well. At later times, however, the free fermion entanglement crosses over to a square root growth, which results from the fact that the particle transport is diffusive in that system and that entanglement is produced by particle propagation. By contrast, for the Heisenberg chain, interactions become relevant at that time scale, and the entanglement growth departs from the free fermion result. For times $tJ \gtrsim 1/2$, there is a crossover regime to the late-time KPZ scaling, which explains the deviations of the KPZ scaling for short times in \fig{fig:entanglement} of the main text.

 \section{Strong noise expansion \label{app:C}}
 
 \textbf{The effective Lindblad operator for $\hat \rho$.---}In this section we construct the effective Lindblad superoperator in the strong-noise limit~\cite{cai_algebraic_2013}. We separate the Lindblad superoperator into a dissipative part $\mathcal{{L}}_0 \hat \rho = \sum_j [\hat S_j^z \hat \rho \hat  S_j^z - \frac14 \hat \rho]$, which dominates for strong noise, and a perturbative, coherent part $\mathcal{{L}}_1 \hat \rho = -i [\hat H, \hat \rho]$. We construct the effective Lindblad operator by second order perturbation theory $\mathcal{{L}}_\text{eff}^\rho = \mathcal{\hat{P}} \mathcal{{L}}_1 \frac{1}{\lambda_0 - \mathcal{{L}}_0} \mathcal{{L}}_1 \mathcal{\hat{P}}$, \eq{eq:Leff}.
 
 The steady states of the unperturbed term $\mathcal{{L}}_0$ are spin configurations in the z-basis, $\ket{\ve{s}}\bra{\ve{s}}$, as $\mathcal{{L}}_0 \ket{\ve{s}}\bra{\ve{s}} = \Lambda^2 \sum_j \hat S_j^z \ket{\ve{s}}\bra{\ve{s}} \hat S_j^z - \frac{1}{4} \ket{\ve{s}}\bra{\ve{s}} = 0$. The perturbation $\mathcal{{L}}_1$ generates the coherent time evolution with the Heisenberg Hamiltonian with next-to-nearest neighbor spin exchange $\hat H = -J \sum_j [ \frac12 (S^+_j S^-_{j+1}+S^-_j S^{+}_{j+1}) +\frac{\delta}{2} (S^+_j S^-_{j+2}+S^-_j S^{+}_{j+2}) + S^z_j S^z_j ]$. Since the ferromagnetic coupling commutes with  $\ket{\ve{s}}\bra{\ve{s}}$, it does not generate dynamics. The action of $\mathcal{L}_0$ on the operator product $S^+_j S^-_{j+a}$ is \
 \begin{eqnarray*}
 	&&\mathcal{L}_0(S^+_j S^-_{j+a}) = \Lambda^2 \sum_i(S_i^z S^+_j S^-_{j+a} S_i^z - \frac{1}{4} S^+_j S^-_{j+a}) \\&&= \Lambda^2 (S_j^z S^+_j S_j^z S^-_{j+a} + S^+_j S_{j+a}^z S^-_{j+a} S_{j+a}^z -  \frac{1}{2} S^+_j S^-_{j+a}) \\&&=  -\Lambda^2 S_j^+ S_{j+a}^-
 \end{eqnarray*}
 and similarly
 $$\mathcal{L}_0(S^-_j S^+_{j+a})  = -\Lambda^2 S_j^- S_{j+a}^+$$
 Therefore, we obtain for the effective Lindblad operator~\cite{cai_algebraic_2013}
 \begin{eqnarray}
 \mathcal{{L}}_\text{eff}^\rho  = \frac{1}{\Lambda^2} \mathcal{\hat{P}} (\mathcal{{L}}_1)^2 \mathcal{\hat{P}}. 
 \label{eq:sLeff}
 \end{eqnarray}

 \textbf{The effective Lindblad operator for $\hat \rho \otimes \hat \rho$.---}The purity can be calculated by judiciously contracting the indices of $\hat \rho \otimes \hat \rho$. The OTOC can be obtained in a similar way by time evolving $S^z_i(t) \otimes S^z_i(t)$ and contracting the initial state with $S^z_0(0)$. Therefore, we will analyze the strong coupling limit of the Lindblad equation for $\hat \rho \otimes \hat \rho$, \eq{eq:LL},
 \begin{eqnarray*}
 	&&\frac{\partial (\hat \rho\otimes \hat \rho)}{\partial t} = -i [\mathcal{\hat H}, \hat \rho\otimes \hat \rho] + \\&&{\Lambda^2} \sum_j \Big(\mathcal{\hat S}_j^z \hat \rho\otimes \hat \rho \mathcal{\hat S}_j^z \nonumber-  \frac12 [(\mathcal{\hat S}_j^z)^2 \hat \rho\otimes \hat \rho - \hat \rho\otimes \hat \rho (\mathcal{\hat S}_j^z)^2] \Big).
 \end{eqnarray*}
 Here, $\mathcal{\hat H} = \hat H \otimes \identity + \identity \otimes \hat H$ and $\mathcal{\hat S}_j^z = \hat S_j^z \otimes \identity + \identity \otimes \hat S_j^z$ act on the product space. We proceed now similarly to the construction of $\mathcal{L}_\text{eff}$ for the density matrix by introducing the unperturbed Lindblad operator $\mathcal{L}_0 = {\Lambda^2} \sum_j \Big(\mathcal{\hat S}_j^z \hat \rho\otimes \hat \rho \mathcal{\hat S}_j^z \nonumber-  \frac12 [(\mathcal{\hat S}_j^z)^2 \hat \rho\otimes \hat \rho - \hat \rho\otimes \hat \rho (\mathcal{\hat S}_j^z)^2]\Big)$ and the perturbed one $\mathcal{L}_1 = -i [\mathcal{\hat H}, \hat \rho\otimes \hat \rho]$.
 
 First, we find the steady states of $\mathcal{L}_0$. By noting that $(\mathcal{\hat S}_j^z)^2 = 2[\hat S_j^z \otimes\hat S_j^z +\frac14]$, we can read off the following classes of steady states (see also Ref.~\onlinecite{rowlands_noisy_2018}), $\ket{\ve s, \ve u}\bra{\ve s, \ve u}$ and $\ket{\ve s, \bar{\ve s}}\bra{\ve s, \bar{\ve s}}$, where $\ve s$, $\ve u$ are configurations of spin-z eigenstates and $\bar{\ve s}$ is the spin flipped configuration of $\ve{s}$. Let us calculate the effect of $\mathcal{L}_0$ on the perturbed states $\mathcal{L}_1 \ket{\ve s, \ve u}\bra{\ve s', \ve u'} $:
 \begin{eqnarray}
 &&\mathcal{L}_0 (\mathcal{L}_1 \ket{\ve s, \ve u}\bra{\ve s', \ve u'}) = {\Lambda^2} \sum_j \Big(\mathcal{\hat S}_j^z \mathcal{L}_1 \ket{\ve s, \ve u}\bra{\ve s', \ve u'} \mathcal{\hat S}_j^z \nonumber \\&& -\frac12 [(\mathcal{\hat S}_j^z)^2 \mathcal{L}_1 \ket{\ve s, \ve u}\bra{\ve s', \ve u'} - \mathcal{L}_1 \ket{\ve s, \ve u}\bra{\ve s', \ve u'} (\mathcal{\hat S}_j^z)^2]\Big)
 \end{eqnarray}
 Evaluating this expression for the steady states, $\ket{\ve s, \ve u}\bra{\ve s, \ve u}$ and $\ket{\ve s, \bar{\ve s}}\bra{\bar{\ve s}, \ve s}$ similarly as before, we find that in both cases they have an eigenvalue $\Lambda^2$. Hence, we find in total 
 \begin{eqnarray}
 \mathcal{L}_\text{eff}^{\rho\otimes\rho} = -\frac{1}{\Lambda^2}  \mathcal{\hat{P}} (\mathcal{{L}}_1)^2 \mathcal{\hat{P}} .
 \label{eq:Leffrhorho}
 \end{eqnarray}
 The scaling with noise is the same as for conventional time ordered expectation values,  \eq{eq:sLeff}. As a consequence, transport, operator scrambling, and  entanglement production is in the strong noise limit governed by the same effective  scaling with the noise strength, \textit{cf.} \eq{eq:timescale},
 \begin{equation*}
 \tau^{-1} \sim (1+\delta)^2 \frac{J^2}{\Lambda^2},
 \end{equation*}
 where the factor $(1+\delta)^2 J^2$ comes from the nearest and next-to-nearest neighbor in-plane Heisenberg coupling in $\mathcal{L}_1$, which is applied twice in \eq{eq:Leffrhorho}.

\begin{acknowledgments}
	M.K. thanks Sarang Gopalakrishnan, Adam Nahum, Frank Pollmann, Tibor Rakovszky, Matteo Scandi, and Herbert Spohn for interesting discussions. This work was supported by the Technical University of Munich - Institute for Advanced Study, funded by the German Excellence Initiative and the European Union FP7 under grant agreement 291763, by the DFG grant No. KN 1254/1-1, and DFG TRR80 (Project F8).
\end{acknowledgments}


\begin{thebibliography}{62}%
	\makeatletter
	\providecommand \@ifxundefined [1]{%
		\@ifx{#1\undefined}
	}%
	\providecommand \@ifnum [1]{%
		\ifnum #1\expandafter \@firstoftwo
		\else \expandafter \@secondoftwo
		\fi
	}%
	\providecommand \@ifx [1]{%
		\ifx #1\expandafter \@firstoftwo
		\else \expandafter \@secondoftwo
		\fi
	}%
	\providecommand \natexlab [1]{#1}%
	\providecommand \enquote  [1]{``#1''}%
	\providecommand \bibnamefont  [1]{#1}%
	\providecommand \bibfnamefont [1]{#1}%
	\providecommand \citenamefont [1]{#1}%
	\providecommand \href@noop [0]{\@secondoftwo}%
	\providecommand \href [0]{\begingroup \@sanitize@url \@href}%
	\providecommand \@href[1]{\@@startlink{#1}\@@href}%
	\providecommand \@@href[1]{\endgroup#1\@@endlink}%
	\providecommand \@sanitize@url [0]{\catcode `\\12\catcode `\$12\catcode
		`\&12\catcode `\#12\catcode `\^12\catcode `\_12\catcode `\%12\relax}%
	\providecommand \@@startlink[1]{}%
	\providecommand \@@endlink[0]{}%
	\providecommand \url  [0]{\begingroup\@sanitize@url \@url }%
	\providecommand \@url [1]{\endgroup\@href {#1}{\urlprefix }}%
	\providecommand \urlprefix  [0]{URL }%
	\providecommand \Eprint [0]{\href }%
	\providecommand \doibase [0]{http://dx.doi.org/}%
	\providecommand \selectlanguage [0]{\@gobble}%
	\providecommand \bibinfo  [0]{\@secondoftwo}%
	\providecommand \bibfield  [0]{\@secondoftwo}%
	\providecommand \translation [1]{[#1]}%
	\providecommand \BibitemOpen [0]{}%
	\providecommand \bibitemStop [0]{}%
	\providecommand \bibitemNoStop [0]{.\EOS\space}%
	\providecommand \EOS [0]{\spacefactor3000\relax}%
	\providecommand \BibitemShut  [1]{\csname bibitem#1\endcsname}%
	\let\auto@bib@innerbib\@empty
	\bibitem [{\citenamefont {Deutsch}(1991)}]{deutsch_quantum_1991}%
	\BibitemOpen
	\bibfield  {author} {\bibinfo {author} {\bibfnamefont {J.~M.}\ \bibnamefont
			{Deutsch}},\ }\bibfield  {title} {\enquote {\bibinfo {title} {Quantum
				statistical mechanics in a closed system},}\ }\href {\doibase
		10.1103/PhysRevA.43.2046} {\bibfield  {journal} {\bibinfo  {journal} {Phys.
				Rev. A}\ }\textbf {\bibinfo {volume} {43}},\ \bibinfo {pages} {2046--2049}
		(\bibinfo {year} {1991})}\BibitemShut {NoStop}%
	\bibitem [{\citenamefont {Srednicki}(1994)}]{srednicki_chaos_1994}%
	\BibitemOpen
	\bibfield  {author} {\bibinfo {author} {\bibfnamefont {Mark}\ \bibnamefont
			{Srednicki}},\ }\bibfield  {title} {\enquote {\bibinfo {title} {Chaos and
				quantum thermalization},}\ }\href {\doibase 10.1103/PhysRevE.50.888}
	{\bibfield  {journal} {\bibinfo  {journal} {Phys. Rev. E}\ }\textbf {\bibinfo
			{volume} {50}},\ \bibinfo {pages} {888--901} (\bibinfo {year}
		{1994})}\BibitemShut {NoStop}%
	\bibitem [{\citenamefont {Rigol}\ \emph {et~al.}(2008)\citenamefont {Rigol},
		\citenamefont {Dunjko},\ and\ \citenamefont
		{Olshanii}}]{rigol_thermalization_2008}%
	\BibitemOpen
	\bibfield  {author} {\bibinfo {author} {\bibfnamefont {Marcos}\ \bibnamefont
			{Rigol}}, \bibinfo {author} {\bibfnamefont {Vanja}\ \bibnamefont {Dunjko}}, \
		and\ \bibinfo {author} {\bibfnamefont {Maxim}\ \bibnamefont {Olshanii}},\
	}\bibfield  {title} {\enquote {\bibinfo {title} {Thermalization and its
			mechanism for generic isolated quantum systems},}\ }\href {\doibase
	10.1038/nature06838} {\bibfield  {journal} {\bibinfo  {journal} {Nature
			(London)}\ }\textbf {\bibinfo {volume} {452}},\ \bibinfo {pages} {854--858}
	(\bibinfo {year} {2008})}\BibitemShut {NoStop}%
\bibitem [{\citenamefont {Calabrese}\ and\ \citenamefont
	{Cardy}(2005)}]{calabrese_evolution_2005}%
\BibitemOpen
\bibfield  {author} {\bibinfo {author} {\bibfnamefont {Pasquale}\
		\bibnamefont {Calabrese}}\ and\ \bibinfo {author} {\bibfnamefont {John}\
		\bibnamefont {Cardy}},\ }\bibfield  {title} {\enquote {\bibinfo {title}
		{Evolution of entanglement entropy in one-dimensional systems},}\ }\href
{\doibase 10.1088/1742-5468/2005/04/P04010} {\bibfield  {journal} {\bibinfo
		{journal} {J. Stat. Mech.}\ }\textbf {\bibinfo {volume} {2005}},\ \bibinfo
	{pages} {P04010} (\bibinfo {year} {2005})}\BibitemShut {NoStop}%
\bibitem [{\citenamefont {Kim}\ and\ \citenamefont
	{Huse}(2013)}]{kim_ballistic_2013}%
\BibitemOpen
\bibfield  {author} {\bibinfo {author} {\bibfnamefont {Hyungwon}\
		\bibnamefont {Kim}}\ and\ \bibinfo {author} {\bibfnamefont {David~A.}\
		\bibnamefont {Huse}},\ }\bibfield  {title} {\enquote {\bibinfo {title}
		{Ballistic {Spreading} of {Entanglement} in a {Diffusive} {Nonintegrable}
			{System}},}\ }\href {\doibase 10.1103/PhysRevLett.111.127205} {\bibfield
	{journal} {\bibinfo  {journal} {Phys. Rev. Lett.}\ }\textbf {\bibinfo
		{volume} {111}},\ \bibinfo {pages} {127205} (\bibinfo {year}
	{2013})}\BibitemShut {NoStop}%
\bibitem [{\citenamefont {Mezei}\ and\ \citenamefont
	{Stanford}(2017)}]{mezei_entanglement_2017-1}%
\BibitemOpen
\bibfield  {author} {\bibinfo {author} {\bibfnamefont {Márk}\ \bibnamefont
		{Mezei}}\ and\ \bibinfo {author} {\bibfnamefont {Douglas}\ \bibnamefont
		{Stanford}},\ }\bibfield  {title} {\enquote {\bibinfo {title} {On
			entanglement spreading in chaotic systems},}\ }\href {\doibase
	10.1007/JHEP05(2017)065} {\bibfield  {journal} {\bibinfo  {journal} {J. High
			Energ. Phys.}\ }\textbf {\bibinfo {volume} {2017}},\ \bibinfo {pages} {65}
	(\bibinfo {year} {2017})}\BibitemShut {NoStop}%
\bibitem [{\citenamefont {Nahum}\ \emph {et~al.}(2017)\citenamefont {Nahum},
	\citenamefont {Ruhman}, \citenamefont {Vijay},\ and\ \citenamefont
	{Haah}}]{nahum_quantum_2017}%
\BibitemOpen
\bibfield  {author} {\bibinfo {author} {\bibfnamefont {Adam}\ \bibnamefont
		{Nahum}}, \bibinfo {author} {\bibfnamefont {Jonathan}\ \bibnamefont
		{Ruhman}}, \bibinfo {author} {\bibfnamefont {Sagar}\ \bibnamefont {Vijay}}, \
	and\ \bibinfo {author} {\bibfnamefont {Jeongwan}\ \bibnamefont {Haah}},\
}\bibfield  {title} {\enquote {\bibinfo {title} {Quantum entanglement growth
		under random unitary dynamics},}\ }\href {\doibase 10.1103/PhysRevX.7.031016}
{\bibfield  {journal} {\bibinfo  {journal} {Phys. Rev. X}\ }\textbf {\bibinfo
		{volume} {7}},\ \bibinfo {pages} {031016} (\bibinfo {year}
	{2017})}\BibitemShut {NoStop}%
\bibitem [{\citenamefont {Zhou}\ and\ \citenamefont
	{Nahum}(2018)}]{zhou_emergent_2018}%
\BibitemOpen
\bibfield  {author} {\bibinfo {author} {\bibfnamefont {Tianci}\ \bibnamefont
		{Zhou}}\ and\ \bibinfo {author} {\bibfnamefont {Adam}\ \bibnamefont
		{Nahum}},\ }\bibfield  {title} {\enquote {\bibinfo {title} {Emergent
			statistical mechanics of entanglement in random unitary circuits},}\ }\href
{http://arxiv.org/abs/1804.09737} {\bibfield  {journal} {\bibinfo  {journal}
		{{arXiv}:1804.09737}\ } (\bibinfo {year} {2018})}\BibitemShut {NoStop}%
\bibitem [{\citenamefont {A.I.}\ and\ \citenamefont {N.}(1969)}]{larkin_1969}%
\BibitemOpen
\bibfield  {author} {\bibinfo {author} {\bibfnamefont {Larkin}\ \bibnamefont
		{A.I.}}\ and\ \bibinfo {author} {\bibfnamefont {Ovchinnikov~Yu.}\
		\bibnamefont {N.}},\ }\bibfield  {title} {\enquote {\bibinfo {title}
		{Quasiclassical method in the theory of superconductivity.}}\ }\href@noop {}
{\bibfield  {journal} {\bibinfo  {journal} {JETP Lett.}\ }\textbf {\bibinfo
		{volume} {28}},\ \bibinfo {pages} {1200} (\bibinfo {year}
	{1969})}\BibitemShut {NoStop}%
\bibitem [{\citenamefont {Kitaev}(2015)}]{kitaev_2014}%
\BibitemOpen
\bibfield  {author} {\bibinfo {author} {\bibfnamefont {A.~Y.}\ \bibnamefont
		{Kitaev}},\ }\bibfield  {title} {\enquote {\bibinfo {title} {A simple model
			of quantum holography},}\ }\href
{http://online.kitp.ucsb.edu/online/entangled15/} {\bibfield  {journal}
	{\bibinfo  {journal} {KITP Program on Entanglement in Strongly-Correlated
			Quantum Matter}\ } (\bibinfo {year} {2015})}\BibitemShut {NoStop}%
\bibitem [{\citenamefont {Shenker}\ and\ \citenamefont
	{Stanford}(2014)}]{shenker_black_2014}%
\BibitemOpen
\bibfield  {author} {\bibinfo {author} {\bibfnamefont {Stephen~H.}\
		\bibnamefont {Shenker}}\ and\ \bibinfo {author} {\bibfnamefont {Douglas}\
		\bibnamefont {Stanford}},\ }\bibfield  {title} {\enquote {\bibinfo {title}
		{Black holes and the butterfly effect},}\ }\href {\doibase
	10.1007/JHEP03(2014)067} {\bibfield  {journal} {\bibinfo  {journal} {J. High
			Energy Phys.}\ }\textbf {\bibinfo {volume} {2014}},\ \bibinfo {pages} {67}
	(\bibinfo {year} {2014})}\BibitemShut {NoStop}%
\bibitem [{\citenamefont {Hosur}\ \emph {et~al.}(2016)\citenamefont {Hosur},
	\citenamefont {Qi}, \citenamefont {Roberts},\ and\ \citenamefont
	{Yoshida}}]{hosur_chaos_2016}%
\BibitemOpen
\bibfield  {author} {\bibinfo {author} {\bibfnamefont {Pavan}\ \bibnamefont
		{Hosur}}, \bibinfo {author} {\bibfnamefont {Xiao-Liang}\ \bibnamefont {Qi}},
	\bibinfo {author} {\bibfnamefont {Daniel~A.}\ \bibnamefont {Roberts}}, \ and\
	\bibinfo {author} {\bibfnamefont {Beni}\ \bibnamefont {Yoshida}},\ }\bibfield
{title} {\enquote {\bibinfo {title} {Chaos in quantum channels},}\ }\href
{\doibase 10.1007/JHEP02(2016)004} {\bibfield  {journal} {\bibinfo  {journal}
		{J. High Energy Phys.}\ }\textbf {\bibinfo {volume} {2016}},\ \bibinfo
	{pages} {4} (\bibinfo {year} {2016})}\BibitemShut {NoStop}%
\bibitem [{\citenamefont {Polchinski}\ and\ \citenamefont
	{Rosenhaus}(2016)}]{polchinski_spectrum_2016}%
\BibitemOpen
\bibfield  {author} {\bibinfo {author} {\bibfnamefont {Joseph}\ \bibnamefont
		{Polchinski}}\ and\ \bibinfo {author} {\bibfnamefont {Vladimir}\ \bibnamefont
		{Rosenhaus}},\ }\bibfield  {title} {\enquote {\bibinfo {title} {The spectrum
			in the {Sachdev}-{Ye}-{Kitaev} model},}\ }\href {\doibase
	10.1007/JHEP04(2016)001} {\bibfield  {journal} {\bibinfo  {journal} {J. High
			Energy Phys.}\ }\textbf {\bibinfo {volume} {2016}},\ \bibinfo {pages} {1}
	(\bibinfo {year} {2016})}\BibitemShut {NoStop}%
\bibitem [{\citenamefont {Maldacena}\ \emph {et~al.}(2016)\citenamefont
	{Maldacena}, \citenamefont {Shenker},\ and\ \citenamefont
	{Stanford}}]{maldacena_bound_2016}%
\BibitemOpen
\bibfield  {author} {\bibinfo {author} {\bibfnamefont {Juan}\ \bibnamefont
		{Maldacena}}, \bibinfo {author} {\bibfnamefont {Stephen~H.}\ \bibnamefont
		{Shenker}}, \ and\ \bibinfo {author} {\bibfnamefont {Douglas}\ \bibnamefont
		{Stanford}},\ }\bibfield  {title} {\enquote {\bibinfo {title} {A bound on
			chaos},}\ }\href {\doibase 10.1007/JHEP08(2016)106} {\bibfield  {journal}
	{\bibinfo  {journal} {J. High Energy Phys.}\ }\textbf {\bibinfo {volume}
		{2016}},\ \bibinfo {pages} {106} (\bibinfo {year} {2016})}\BibitemShut
{NoStop}%
\bibitem [{\citenamefont {Roberts}\ and\ \citenamefont
	{Stanford}(2015)}]{roberts_diagnosing_2015}%
\BibitemOpen
\bibfield  {author} {\bibinfo {author} {\bibfnamefont {Daniel~A.}\
		\bibnamefont {Roberts}}\ and\ \bibinfo {author} {\bibfnamefont {Douglas}\
		\bibnamefont {Stanford}},\ }\bibfield  {title} {\enquote {\bibinfo {title}
		{Diagnosing chaos using four-point functions in two-dimensional conformal
			field theory},}\ }\href {\doibase 10.1103/PhysRevLett.115.131603} {\bibfield
	{journal} {\bibinfo  {journal} {Phys. Rev. Lett.}\ }\textbf {\bibinfo
		{volume} {115}},\ \bibinfo {pages} {131603} (\bibinfo {year}
	{2015})}\BibitemShut {NoStop}%
\bibitem [{\citenamefont {Stanford}(2016)}]{stanford_many-body_2016}%
\BibitemOpen
\bibfield  {author} {\bibinfo {author} {\bibfnamefont {Douglas}\ \bibnamefont
		{Stanford}},\ }\bibfield  {title} {\enquote {\bibinfo {title} {Many-body
			chaos at weak coupling},}\ }\href {\doibase 10.1007/JHEP10(2016)009}
{\bibfield  {journal} {\bibinfo  {journal} {J. High Energy Phys.}\ }\textbf
	{\bibinfo {volume} {2016}},\ \bibinfo {pages} {9} (\bibinfo {year}
	{2016})}\BibitemShut {NoStop}%
\bibitem [{\citenamefont {Patel}\ \emph {et~al.}(2017)\citenamefont {Patel},
	\citenamefont {Chowdhury}, \citenamefont {Sachdev},\ and\ \citenamefont
	{Swingle}}]{patel_quantum_2017}%
\BibitemOpen
\bibfield  {author} {\bibinfo {author} {\bibfnamefont {Aavishkar~A.}\
		\bibnamefont {Patel}}, \bibinfo {author} {\bibfnamefont {Debanjan}\
		\bibnamefont {Chowdhury}}, \bibinfo {author} {\bibfnamefont {Subir}\
		\bibnamefont {Sachdev}}, \ and\ \bibinfo {author} {\bibfnamefont {Brian}\
		\bibnamefont {Swingle}},\ }\bibfield  {title} {\enquote {\bibinfo {title}
		{Quantum butterfly effect in weakly interacting diffusive metals},}\ }\href
{\doibase 10.1103/PhysRevX.7.031047} {\bibfield  {journal} {\bibinfo
		{journal} {Phys. Rev. X}\ }\textbf {\bibinfo {volume} {7}},\ \bibinfo {pages}
	{031047} (\bibinfo {year} {2017})}\BibitemShut {NoStop}%
\bibitem [{\citenamefont {Werman}\ \emph {et~al.}(2017)\citenamefont {Werman},
	\citenamefont {Kivelson},\ and\ \citenamefont {Berg}}]{werman_quantum_2017}%
\BibitemOpen
\bibfield  {author} {\bibinfo {author} {\bibfnamefont {Yochai}\ \bibnamefont
		{Werman}}, \bibinfo {author} {\bibfnamefont {Steven~A.}\ \bibnamefont
		{Kivelson}}, \ and\ \bibinfo {author} {\bibfnamefont {Erez}\ \bibnamefont
		{Berg}},\ }\bibfield  {title} {\enquote {\bibinfo {title} {Quantum chaos in
			an electron-phonon bad metal},}\ }\href {http://arxiv.org/abs/1705.07895}
{\bibfield  {journal} {\bibinfo  {journal} {{arXiv}:1705.07895}\ } (\bibinfo
	{year} {2017})}\BibitemShut {NoStop}%
\bibitem [{\citenamefont {Gu}\ \emph {et~al.}(2017)\citenamefont {Gu},
	\citenamefont {Qi},\ and\ \citenamefont {Stanford}}]{gu_local_2017}%
\BibitemOpen
\bibfield  {author} {\bibinfo {author} {\bibfnamefont {Yingfei}\ \bibnamefont
		{Gu}}, \bibinfo {author} {\bibfnamefont {Xiao-Liang}\ \bibnamefont {Qi}}, \
	and\ \bibinfo {author} {\bibfnamefont {Douglas}\ \bibnamefont {Stanford}},\
}\bibfield  {title} {\enquote {\bibinfo {title} {Local criticality, diffusion
		and chaos in generalized sachdev-ye-kitaev models},}\ }\href {\doibase
10.1007/JHEP05(2017)125} {\bibfield  {journal} {\bibinfo  {journal} {J. High
		Energ. Phys.}\ }\textbf {\bibinfo {volume} {2017}},\ \bibinfo {pages} {125}
(\bibinfo {year} {2017})}\BibitemShut {NoStop}%
\bibitem [{\citenamefont {Jian}\ and\ \citenamefont
	{Yao}(2018)}]{jian_universal_2018}%
\BibitemOpen
\bibfield  {author} {\bibinfo {author} {\bibfnamefont {Shao-Kai}\
		\bibnamefont {Jian}}\ and\ \bibinfo {author} {\bibfnamefont {Hong}\
		\bibnamefont {Yao}},\ }\bibfield  {title} {\enquote {\bibinfo {title}
		{Universal properties of many-body quantum chaos at gross-neveu
			criticality},}\ }\href {http://arxiv.org/abs/1805.12299} {\bibfield
	{journal} {\bibinfo  {journal} {{arXiv}:1805.12299}\ } (\bibinfo {year}
	{2018})}\BibitemShut {NoStop}%
\bibitem [{\citenamefont {Aleiner}\ \emph {et~al.}(2016)\citenamefont
	{Aleiner}, \citenamefont {Faoro},\ and\ \citenamefont
	{Ioffe}}]{aleiner_microscopic_2016}%
\BibitemOpen
\bibfield  {author} {\bibinfo {author} {\bibfnamefont {Igor~L.}\ \bibnamefont
		{Aleiner}}, \bibinfo {author} {\bibfnamefont {Lara}\ \bibnamefont {Faoro}}, \
	and\ \bibinfo {author} {\bibfnamefont {Lev~B.}\ \bibnamefont {Ioffe}},\
}\bibfield  {title} {\enquote {\bibinfo {title} {Microscopic model of quantum
		butterfly effect: Out-of-time-order correlators and traveling combustion
		waves},}\ }\href {\doibase 10.1016/j.aop.2016.09.006} {\bibfield  {journal}
{\bibinfo  {journal} {Annals of Physics}\ }\textbf {\bibinfo {volume}
	{375}},\ \bibinfo {pages} {378--406} (\bibinfo {year} {2016})}\BibitemShut
{NoStop}%
\bibitem [{\citenamefont {Kurchan}(2016)}]{kurchan_quantum_2016}%
\BibitemOpen
\bibfield  {author} {\bibinfo {author} {\bibfnamefont {Jorge}\ \bibnamefont
		{Kurchan}},\ }\bibfield  {title} {\enquote {\bibinfo {title} {Quantum bound
			to chaos and the semiclassical limit},}\ }\href
{http://arxiv.org/abs/1612.01278} {\bibfield  {journal} {\bibinfo  {journal}
		{{arXiv}:1612.01278}\ } (\bibinfo {year} {2016})}\BibitemShut {NoStop}%
\bibitem [{\citenamefont {Rozenbaum}\ \emph {et~al.}(2017)\citenamefont
	{Rozenbaum}, \citenamefont {Ganeshan},\ and\ \citenamefont
	{Galitski}}]{rozenbaum_lyapunov_2017}%
\BibitemOpen
\bibfield  {author} {\bibinfo {author} {\bibfnamefont {Efim~B.}\ \bibnamefont
		{Rozenbaum}}, \bibinfo {author} {\bibfnamefont {Sriram}\ \bibnamefont
		{Ganeshan}}, \ and\ \bibinfo {author} {\bibfnamefont {Victor}\ \bibnamefont
		{Galitski}},\ }\bibfield  {title} {\enquote {\bibinfo {title} {Lyapunov
			exponent and out-of-time-ordered correlator's growth rate in a chaotic
			system},}\ }\href {\doibase 10.1103/PhysRevLett.118.086801} {\bibfield
	{journal} {\bibinfo  {journal} {Phys. Rev. Lett.}\ }\textbf {\bibinfo
		{volume} {118}},\ \bibinfo {pages} {086801} (\bibinfo {year}
	{2017})}\BibitemShut {NoStop}%
\bibitem [{\citenamefont {Rammensee}\ \emph {et~al.}(2018)\citenamefont
	{Rammensee}, \citenamefont {Urbina},\ and\ \citenamefont
	{Richter}}]{rammensee_many-body_2018}%
\BibitemOpen
\bibfield  {author} {\bibinfo {author} {\bibfnamefont {Josef}\ \bibnamefont
		{Rammensee}}, \bibinfo {author} {\bibfnamefont {Juan~Diego}\ \bibnamefont
		{Urbina}}, \ and\ \bibinfo {author} {\bibfnamefont {Klaus}\ \bibnamefont
		{Richter}},\ }\bibfield  {title} {\enquote {\bibinfo {title} {Many-body
			quantum interference and the saturation of out-of-time-order correlators},}\
}\href {\doibase 10.1103/PhysRevLett.121.124101} {\bibfield  {journal}
{\bibinfo  {journal} {Phys. Rev. Lett.}\ }\textbf {\bibinfo {volume} {121}},\
\bibinfo {pages} {124101} (\bibinfo {year} {2018})}\BibitemShut {NoStop}%
\bibitem [{\citenamefont {Khemani}\ \emph
	{et~al.}(2018{\natexlab{a}})\citenamefont {Khemani}, \citenamefont {Huse},\
	and\ \citenamefont {Nahum}}]{khemani_velocity-dependent_2018}%
\BibitemOpen
\bibfield  {author} {\bibinfo {author} {\bibfnamefont {Vedika}\ \bibnamefont
		{Khemani}}, \bibinfo {author} {\bibfnamefont {David~A.}\ \bibnamefont
		{Huse}}, \ and\ \bibinfo {author} {\bibfnamefont {Adam}\ \bibnamefont
		{Nahum}},\ }\bibfield  {title} {\enquote {\bibinfo {title}
		{Velocity-dependent lyapunov exponents in many-body quantum, semiclassical,
			and classical chaos},}\ }\href {\doibase 10.1103/PhysRevB.98.144304}
{\bibfield  {journal} {\bibinfo  {journal} {Phys. Rev. B}\ }\textbf {\bibinfo
		{volume} {98}},\ \bibinfo {pages} {144304} (\bibinfo {year}
	{2018}{\natexlab{a}})}\BibitemShut {NoStop}%
\bibitem [{\citenamefont {Bohrdt}\ \emph {et~al.}(2017)\citenamefont {Bohrdt},
	\citenamefont {Mendl}, \citenamefont {Endres},\ and\ \citenamefont
	{Knap}}]{bohrdt_scrambling_2017}%
\BibitemOpen
\bibfield  {author} {\bibinfo {author} {\bibfnamefont {A.}~\bibnamefont
		{Bohrdt}}, \bibinfo {author} {\bibfnamefont {C.~B.}\ \bibnamefont {Mendl}},
	\bibinfo {author} {\bibfnamefont {M.}~\bibnamefont {Endres}}, \ and\ \bibinfo
	{author} {\bibfnamefont {M.}~\bibnamefont {Knap}},\ }\bibfield  {title}
{\enquote {\bibinfo {title} {Scrambling and thermalization in a diffusive
			quantum many-body system},}\ }\href {\doibase 10.1088/1367-2630/aa719b}
{\bibfield  {journal} {\bibinfo  {journal} {New J. Phys.}\ }\textbf {\bibinfo
		{volume} {19}},\ \bibinfo {pages} {063001} (\bibinfo {year}
	{2017})}\BibitemShut {NoStop}%
\bibitem [{\citenamefont {Chen}\ \emph {et~al.}(2017)\citenamefont {Chen},
	\citenamefont {Zhou}, \citenamefont {Huse},\ and\ \citenamefont
	{Fradkin}}]{chen_out--time-order_2017}%
\BibitemOpen
\bibfield  {author} {\bibinfo {author} {\bibfnamefont {Xiao}\ \bibnamefont
		{Chen}}, \bibinfo {author} {\bibfnamefont {Tianci}\ \bibnamefont {Zhou}},
	\bibinfo {author} {\bibfnamefont {David~A.}\ \bibnamefont {Huse}}, \ and\
	\bibinfo {author} {\bibfnamefont {Eduardo}\ \bibnamefont {Fradkin}},\
}\bibfield  {title} {\enquote {\bibinfo {title} {Out-of-time-order
		correlations in many-body localized and thermal phases},}\ }\href {\doibase
10.1002/andp.201600332} {\bibfield  {journal} {\bibinfo  {journal} {Annalen
		der Physik}\ }\textbf {\bibinfo {volume} {529}},\ \bibinfo {pages} {1600332}
(\bibinfo {year} {2017})}\BibitemShut {NoStop}%
\bibitem [{\citenamefont {Luitz}\ and\ \citenamefont
	{Bar~Lev}(2017)}]{luitz_information_2017}%
\BibitemOpen
\bibfield  {author} {\bibinfo {author} {\bibfnamefont {David~J.}\
		\bibnamefont {Luitz}}\ and\ \bibinfo {author} {\bibfnamefont {Yevgeny}\
		\bibnamefont {Bar~Lev}},\ }\bibfield  {title} {\enquote {\bibinfo {title}
		{Information propagation in isolated quantum systems},}\ }\href {\doibase
	10.1103/PhysRevB.96.020406} {\bibfield  {journal} {\bibinfo  {journal} {Phys.
			Rev. B}\ }\textbf {\bibinfo {volume} {96}},\ \bibinfo {pages} {020406}
	(\bibinfo {year} {2017})}\BibitemShut {NoStop}%
\bibitem [{\citenamefont {Kukuljan}\ \emph {et~al.}(2017-08-14)\citenamefont
	{Kukuljan}, \citenamefont {Grozdanov},\ and\ \citenamefont
	{Prosen}}]{kukuljan_weak_2017}%
\BibitemOpen
\bibfield  {author} {\bibinfo {author} {\bibfnamefont {Ivan}\ \bibnamefont
		{Kukuljan}}, \bibinfo {author} {\bibfnamefont {Sašo}\ \bibnamefont
		{Grozdanov}}, \ and\ \bibinfo {author} {\bibfnamefont {Tomaž}\ \bibnamefont
		{Prosen}},\ }\bibfield  {title} {\enquote {\bibinfo {title} {Weak quantum
			chaos},}\ }\href {\doibase 10.1103/PhysRevB.96.060301} {\bibfield  {journal}
	{\bibinfo  {journal} {Phys. Rev. B}\ }\textbf {\bibinfo {volume} {96}},\
	\bibinfo {pages} {060301} (\bibinfo {year} {2017-08-14})}\BibitemShut
{NoStop}%
\bibitem [{\citenamefont {Jonay}\ \emph {et~al.}(2018)\citenamefont {Jonay},
	\citenamefont {Huse},\ and\ \citenamefont
	{Nahum}}]{jonay_coarse-grained_2018}%
\BibitemOpen
\bibfield  {author} {\bibinfo {author} {\bibfnamefont {Cheryne}\ \bibnamefont
		{Jonay}}, \bibinfo {author} {\bibfnamefont {David~A.}\ \bibnamefont {Huse}},
	\ and\ \bibinfo {author} {\bibfnamefont {Adam}\ \bibnamefont {Nahum}},\
}\bibfield  {title} {\enquote {\bibinfo {title} {Coarse-grained dynamics of
		operator and state entanglement},}\ }\href {http://arxiv.org/abs/1803.00089}
{\bibfield  {journal} {\bibinfo  {journal} {{arXiv}:1803.00089}\ } (\bibinfo
	{year} {2018})}\BibitemShut {NoStop}%
\bibitem [{\citenamefont {Pappalardi}\ \emph {et~al.}(2018)\citenamefont
	{Pappalardi}, \citenamefont {Russomanno}, \citenamefont {\ifmmode
		\check{Z}\else \v{Z}\fi{}unkovi\ifmmode~\check{c}\else \v{c}\fi{}},
	\citenamefont {Iemini}, \citenamefont {Silva},\ and\ \citenamefont
	{Fazio}}]{pappalardi_scrambling_2018}%
\BibitemOpen
\bibfield  {author} {\bibinfo {author} {\bibfnamefont {Silvia}\ \bibnamefont
		{Pappalardi}}, \bibinfo {author} {\bibfnamefont {Angelo}\ \bibnamefont
		{Russomanno}}, \bibinfo {author} {\bibfnamefont {Bojan}\ \bibnamefont
		{\ifmmode \check{Z}\else \v{Z}\fi{}unkovi\ifmmode~\check{c}\else
			\v{c}\fi{}}}, \bibinfo {author} {\bibfnamefont {Fernando}\ \bibnamefont
		{Iemini}}, \bibinfo {author} {\bibfnamefont {Alessandro}\ \bibnamefont
		{Silva}}, \ and\ \bibinfo {author} {\bibfnamefont {Rosario}\ \bibnamefont
		{Fazio}},\ }\bibfield  {title} {\enquote {\bibinfo {title} {Scrambling and
			entanglement spreading in long-range spin chains},}\ }\href {\doibase
	10.1103/PhysRevB.98.134303} {\bibfield  {journal} {\bibinfo  {journal} {Phys.
			Rev. B}\ }\textbf {\bibinfo {volume} {98}},\ \bibinfo {pages} {134303}
	(\bibinfo {year} {2018})}\BibitemShut {NoStop}%
\bibitem [{\citenamefont {Islam}\ \emph {et~al.}(2015)\citenamefont {Islam},
	\citenamefont {Ma}, \citenamefont {Preiss}, \citenamefont {Eric~Tai},
	\citenamefont {Lukin}, \citenamefont {Rispoli},\ and\ \citenamefont
	{Greiner}}]{islam_measuring_2015}%
\BibitemOpen
\bibfield  {author} {\bibinfo {author} {\bibfnamefont {Rajibul}\ \bibnamefont
		{Islam}}, \bibinfo {author} {\bibfnamefont {Ruichao}\ \bibnamefont {Ma}},
	\bibinfo {author} {\bibfnamefont {Philipp~M.}\ \bibnamefont {Preiss}},
	\bibinfo {author} {\bibfnamefont {M.}~\bibnamefont {Eric~Tai}}, \bibinfo
	{author} {\bibfnamefont {Alexander}\ \bibnamefont {Lukin}}, \bibinfo {author}
	{\bibfnamefont {Matthew}\ \bibnamefont {Rispoli}}, \ and\ \bibinfo {author}
	{\bibfnamefont {Markus}\ \bibnamefont {Greiner}},\ }\bibfield  {title}
{\enquote {\bibinfo {title} {Measuring entanglement entropy in a quantum
			many-body system},}\ }\href {\doibase 10.1038/nature15750} {\bibfield
	{journal} {\bibinfo  {journal} {Nature}\ }\textbf {\bibinfo {volume} {528}},\
	\bibinfo {pages} {77--83} (\bibinfo {year} {2015})}\BibitemShut {NoStop}%
\bibitem [{\citenamefont {Kaufman}\ \emph {et~al.}(2016)\citenamefont
	{Kaufman}, \citenamefont {Tai}, \citenamefont {Lukin}, \citenamefont
	{Rispoli}, \citenamefont {Schittko}, \citenamefont {Preiss},\ and\
	\citenamefont {Greiner}}]{kaufman_quantum_2016}%
\BibitemOpen
\bibfield  {author} {\bibinfo {author} {\bibfnamefont {Adam~M.}\ \bibnamefont
		{Kaufman}}, \bibinfo {author} {\bibfnamefont {M.~Eric}\ \bibnamefont {Tai}},
	\bibinfo {author} {\bibfnamefont {Alexander}\ \bibnamefont {Lukin}}, \bibinfo
	{author} {\bibfnamefont {Matthew}\ \bibnamefont {Rispoli}}, \bibinfo {author}
	{\bibfnamefont {Robert}\ \bibnamefont {Schittko}}, \bibinfo {author}
	{\bibfnamefont {Philipp~M.}\ \bibnamefont {Preiss}}, \ and\ \bibinfo {author}
	{\bibfnamefont {Markus}\ \bibnamefont {Greiner}},\ }\bibfield  {title}
{\enquote {\bibinfo {title} {Quantum thermalization through entanglement in
			an isolated many-body system},}\ }\href {\doibase 10.1126/science.aaf6725}
{\bibfield  {journal} {\bibinfo  {journal} {Science}\ }\textbf {\bibinfo
		{volume} {353}},\ \bibinfo {pages} {794--800} (\bibinfo {year}
	{2016})}\BibitemShut {NoStop}%
\bibitem [{\citenamefont {G\"arttner}\ \emph {et~al.}(2017)\citenamefont
	{G\"arttner}, \citenamefont {Bohnet}, \citenamefont {Safavi-Naini},
	\citenamefont {Wall}, \citenamefont {Bollinger},\ and\ \citenamefont
	{Rey}}]{garttner_measuring_2017}%
\BibitemOpen
\bibfield  {author} {\bibinfo {author} {\bibfnamefont {Martin}\ \bibnamefont
		{G\"arttner}}, \bibinfo {author} {\bibfnamefont {Justin~G.}\ \bibnamefont
		{Bohnet}}, \bibinfo {author} {\bibfnamefont {Arghavan}\ \bibnamefont
		{Safavi-Naini}}, \bibinfo {author} {\bibfnamefont {Michael~L.}\ \bibnamefont
		{Wall}}, \bibinfo {author} {\bibfnamefont {John~J.}\ \bibnamefont
		{Bollinger}}, \ and\ \bibinfo {author} {\bibfnamefont {Ana~Maria}\
		\bibnamefont {Rey}},\ }\bibfield  {title} {\enquote {\bibinfo {title}
		{Measuring out-of-time-order correlations and multiple quantum spectra in a
			trapped-ion quantum magnet},}\ }\href {\doibase 10.1038/nphys4119} {\bibfield
	{journal} {\bibinfo  {journal} {Nat. Phys.}\ }\textbf {\bibinfo {volume}
		{13}},\ \bibinfo {pages} {781--786} (\bibinfo {year} {2017})}\BibitemShut
{NoStop}%
\bibitem [{\citenamefont {Li}\ \emph {et~al.}(2017)\citenamefont {Li},
	\citenamefont {Fan}, \citenamefont {Wang}, \citenamefont {Ye}, \citenamefont
	{Zeng}, \citenamefont {Zhai}, \citenamefont {Peng},\ and\ \citenamefont
	{Du}}]{li_measuring_2017}%
\BibitemOpen
\bibfield  {author} {\bibinfo {author} {\bibfnamefont {Jun}\ \bibnamefont
		{Li}}, \bibinfo {author} {\bibfnamefont {Ruihua}\ \bibnamefont {Fan}},
	\bibinfo {author} {\bibfnamefont {Hengyan}\ \bibnamefont {Wang}}, \bibinfo
	{author} {\bibfnamefont {Bingtian}\ \bibnamefont {Ye}}, \bibinfo {author}
	{\bibfnamefont {Bei}\ \bibnamefont {Zeng}}, \bibinfo {author} {\bibfnamefont
		{Hui}\ \bibnamefont {Zhai}}, \bibinfo {author} {\bibfnamefont {Xinhua}\
		\bibnamefont {Peng}}, \ and\ \bibinfo {author} {\bibfnamefont {Jiangfeng}\
		\bibnamefont {Du}},\ }\bibfield  {title} {\enquote {\bibinfo {title}
		{Measuring out-of-time-order correlators on a nuclear magnetic resonance
			quantum simulator},}\ }\href {\doibase 10.1103/PhysRevX.7.031011} {\bibfield
	{journal} {\bibinfo  {journal} {Phys. Rev. X}\ }\textbf {\bibinfo {volume}
		{7}},\ \bibinfo {pages} {031011} (\bibinfo {year} {2017})}\BibitemShut
{NoStop}%
\bibitem [{\citenamefont {Landsman}\ \emph {et~al.}(2018)\citenamefont
	{Landsman}, \citenamefont {Figgatt}, \citenamefont {Schuster}, \citenamefont
	{Linke}, \citenamefont {Yoshida}, \citenamefont {Yao},\ and\ \citenamefont
	{Monroe}}]{landsman_verified_2018}%
\BibitemOpen
\bibfield  {author} {\bibinfo {author} {\bibfnamefont {Kevin~A.}\
		\bibnamefont {Landsman}}, \bibinfo {author} {\bibfnamefont {Caroline}\
		\bibnamefont {Figgatt}}, \bibinfo {author} {\bibfnamefont {Thomas}\
		\bibnamefont {Schuster}}, \bibinfo {author} {\bibfnamefont {Norbert~M.}\
		\bibnamefont {Linke}}, \bibinfo {author} {\bibfnamefont {Beni}\ \bibnamefont
		{Yoshida}}, \bibinfo {author} {\bibfnamefont {Norm~Y.}\ \bibnamefont {Yao}},
	\ and\ \bibinfo {author} {\bibfnamefont {Christopher}\ \bibnamefont
		{Monroe}},\ }\bibfield  {title} {\enquote {\bibinfo {title} {Verified quantum
			information scrambling},}\ }\href {http://arxiv.org/abs/1806.02807}
{\bibfield  {journal} {\bibinfo  {journal} {{arXiv}:1806.02807}\ } (\bibinfo
	{year} {2018})}\BibitemShut {NoStop}%
\bibitem [{\citenamefont {Nahum}\ \emph {et~al.}(2018)\citenamefont {Nahum},
	\citenamefont {Vijay},\ and\ \citenamefont {Haah}}]{nahum_operator_2018}%
\BibitemOpen
\bibfield  {author} {\bibinfo {author} {\bibfnamefont {Adam}\ \bibnamefont
		{Nahum}}, \bibinfo {author} {\bibfnamefont {Sagar}\ \bibnamefont {Vijay}}, \
	and\ \bibinfo {author} {\bibfnamefont {Jeongwan}\ \bibnamefont {Haah}},\
}\bibfield  {title} {\enquote {\bibinfo {title} {Operator spreading in random
		unitary circuits},}\ }\href {\doibase 10.1103/PhysRevX.8.021014} {\bibfield
{journal} {\bibinfo  {journal} {Phys. Rev. X}\ }\textbf {\bibinfo {volume}
	{8}},\ \bibinfo {pages} {021014} (\bibinfo {year} {2018})}\BibitemShut
{NoStop}%
\bibitem [{\citenamefont {von Keyserlingk}\ \emph {et~al.}(2018)\citenamefont
	{von Keyserlingk}, \citenamefont {Rakovszky}, \citenamefont {Pollmann},\ and\
	\citenamefont {Sondhi}}]{von_keyserlingk_operator_2018}%
\BibitemOpen
\bibfield  {author} {\bibinfo {author} {\bibfnamefont {C. W.}\ \bibnamefont
		{von Keyserlingk}}, \bibinfo {author} {\bibfnamefont {Tibor}\ \bibnamefont
		{Rakovszky}}, \bibinfo {author} {\bibfnamefont {Frank}\ \bibnamefont
		{Pollmann}}, \ and\ \bibinfo {author} {\bibfnamefont {S. L.}\ \bibnamefont
		{Sondhi}},\ }\bibfield  {title} {\enquote {\bibinfo {title} {Operator
			hydrodynamics, {OTOCs}, and entanglement growth in systems without
			conservation laws},}\ }\href {\doibase 10.1103/PhysRevX.8.021013} {\bibfield
	{journal} {\bibinfo  {journal} {Phys. Rev. X}\ }\textbf {\bibinfo {volume}
		{8}},\ \bibinfo {pages} {021013} (\bibinfo {year} {2018})}\BibitemShut
{NoStop}%
\bibitem [{\citenamefont {Khemani}\ \emph
	{et~al.}(2018{\natexlab{b}})\citenamefont {Khemani}, \citenamefont
	{Vishwanath},\ and\ \citenamefont {Huse}}]{khemani_operator_2018}%
\BibitemOpen
\bibfield  {author} {\bibinfo {author} {\bibfnamefont {Vedika}\ \bibnamefont
		{Khemani}}, \bibinfo {author} {\bibfnamefont {Ashvin}\ \bibnamefont
		{Vishwanath}}, \ and\ \bibinfo {author} {\bibfnamefont {David~A.}\
		\bibnamefont {Huse}},\ }\bibfield  {title} {\enquote {\bibinfo {title}
		{Operator spreading and the emergence of dissipative hydrodynamics under
			unitary evolution with conservation laws},}\ }\href {\doibase
	10.1103/PhysRevX.8.031057} {\bibfield  {journal} {\bibinfo  {journal} {Phys.
			Rev. X}\ }\textbf {\bibinfo {volume} {8}},\ \bibinfo {pages} {031057}
	(\bibinfo {year} {2018}{\natexlab{b}})}\BibitemShut {NoStop}%
\bibitem [{\citenamefont {Rakovszky}\ \emph {et~al.}(2018)\citenamefont
	{Rakovszky}, \citenamefont {Pollmann},\ and\ \citenamefont {von
		Keyserlingk}}]{rakovszky_diffusive_2018}%
\BibitemOpen
\bibfield  {author} {\bibinfo {author} {\bibfnamefont {Tibor}\ \bibnamefont
		{Rakovszky}}, \bibinfo {author} {\bibfnamefont {Frank}\ \bibnamefont
		{Pollmann}}, \ and\ \bibinfo {author} {\bibfnamefont {C. W.}\ \bibnamefont
		{von Keyserlingk}},\ }\bibfield  {title} {\enquote {\bibinfo {title}
		{Diffusive hydrodynamics of out-of-time-ordered correlators with charge
			conservation},}\ }\href {\doibase 10.1103/PhysRevX.8.031058} {\bibfield
	{journal} {\bibinfo  {journal} {Phys. Rev. X}\ }\textbf {\bibinfo {volume}
		{8}},\ \bibinfo {pages} {031058} (\bibinfo {year} {2018})}\BibitemShut
{NoStop}%
\bibitem [{\citenamefont {Chan}\ \emph {et~al.}(2017)\citenamefont {Chan},
	\citenamefont {De~Luca},\ and\ \citenamefont {Chalker}}]{chan_solution_2017}%
\BibitemOpen
\bibfield  {author} {\bibinfo {author} {\bibfnamefont {Amos}\ \bibnamefont
		{Chan}}, \bibinfo {author} {\bibfnamefont {Andrea}\ \bibnamefont {De~Luca}},
	\ and\ \bibinfo {author} {\bibfnamefont {J.~T.}\ \bibnamefont {Chalker}},\
}\bibfield  {title} {\enquote {\bibinfo {title} {Solution of a minimal model
		for many-body quantum chaos},}\ }\href {http://arxiv.org/abs/1712.06836}
{\bibfield  {journal} {\bibinfo  {journal} {{arXiv}:1712.06836}\ } (\bibinfo
	{year} {2017})}\BibitemShut {NoStop}%
\bibitem [{\citenamefont {Gopalakrishnan}()}]{gopalakrishnan_operator_2018}%
\BibitemOpen
\bibfield  {author} {\bibinfo {author} {\bibfnamefont {Sarang}\ \bibnamefont
		{Gopalakrishnan}},\ }\bibfield  {title} {\enquote {\bibinfo {title} {Operator
			growth and eigenstate entanglement in an interacting integrable floquet
			system},}\ }\href {\doibase 10.1103/PhysRevB.98.060302} {\bibfield  {journal}
	{\bibinfo  {journal} {Phys. Rev. B}\ }\textbf {\bibinfo {volume} {98}},\
	\bibinfo {pages} {060302}}\BibitemShut {NoStop}%
\bibitem [{\citenamefont {Hamma}\ \emph {et~al.}(2012)\citenamefont {Hamma},
	\citenamefont {Santra},\ and\ \citenamefont {Zanardi}}]{hamma_quantum_2012}%
\BibitemOpen
\bibfield  {author} {\bibinfo {author} {\bibfnamefont {Alioscia}\
		\bibnamefont {Hamma}}, \bibinfo {author} {\bibfnamefont {Siddhartha}\
		\bibnamefont {Santra}}, \ and\ \bibinfo {author} {\bibfnamefont {Paolo}\
		\bibnamefont {Zanardi}},\ }\bibfield  {title} {\enquote {\bibinfo {title}
		{Quantum entanglement in random physical states},}\ }\href {\doibase
	10.1103/PhysRevLett.109.040502} {\bibfield  {journal} {\bibinfo  {journal}
		{Phys. Rev. Lett.}\ }\textbf {\bibinfo {volume} {109}},\ \bibinfo {pages}
	{040502} (\bibinfo {year} {2012})}\BibitemShut {NoStop}%
\bibitem [{\citenamefont {Brown}\ and\ \citenamefont
	{Fawzi}(2015)}]{brown_decoupling_2015}%
\BibitemOpen
\bibfield  {author} {\bibinfo {author} {\bibfnamefont {Winton}\ \bibnamefont
		{Brown}}\ and\ \bibinfo {author} {\bibfnamefont {Omar}\ \bibnamefont
		{Fawzi}},\ }\bibfield  {title} {\enquote {\bibinfo {title} {Decoupling with
			random quantum circuits},}\ }\href {\doibase 10.1007/s00220-015-2470-1}
{\bibfield  {journal} {\bibinfo  {journal} {Commun. Math. Phys.}\ }\textbf
	{\bibinfo {volume} {340}},\ \bibinfo {pages} {867--900} (\bibinfo {year}
	{2015})}\BibitemShut {NoStop}%
\bibitem [{\citenamefont {Kardar}\ \emph {et~al.}(1986)\citenamefont {Kardar},
	\citenamefont {Parisi},\ and\ \citenamefont {Zhang}}]{kardar_dynamic_1986}%
\BibitemOpen
\bibfield  {author} {\bibinfo {author} {\bibfnamefont {Mehran}\ \bibnamefont
		{Kardar}}, \bibinfo {author} {\bibfnamefont {Giorgio}\ \bibnamefont
		{Parisi}}, \ and\ \bibinfo {author} {\bibfnamefont {Yi-Cheng}\ \bibnamefont
		{Zhang}},\ }\bibfield  {title} {\enquote {\bibinfo {title} {Dynamic scaling
			of growing interfaces},}\ }\href {\doibase 10.1103/PhysRevLett.56.889}
{\bibfield  {journal} {\bibinfo  {journal} {Phys. Rev. Lett.}\ }\textbf
	{\bibinfo {volume} {56}},\ \bibinfo {pages} {889--892} (\bibinfo {year}
	{1986})}\BibitemShut {NoStop}%
\bibitem [{\citenamefont {Xu}\ and\ \citenamefont
	{Swingle}(2018)}]{xu_locality_2018}%
\BibitemOpen
\bibfield  {author} {\bibinfo {author} {\bibfnamefont {Shenglong}\
		\bibnamefont {Xu}}\ and\ \bibinfo {author} {\bibfnamefont {Brian}\
		\bibnamefont {Swingle}},\ }\bibfield  {title} {\enquote {\bibinfo {title}
		{Locality, quantum fluctuations, and scrambling},}\ }\href
{http://arxiv.org/abs/1805.05376} {\bibfield  {journal} {\bibinfo  {journal}
		{{arXiv}:1805.05376}\ } (\bibinfo {year} {2018})}\BibitemShut {NoStop}%
\bibitem [{\citenamefont {Rowlands}\ and\ \citenamefont
	{Lamacraft}(2018)}]{rowlands_noisy_2018}%
\BibitemOpen
\bibfield  {author} {\bibinfo {author} {\bibfnamefont {Daniel~A.}\
		\bibnamefont {Rowlands}}\ and\ \bibinfo {author} {\bibfnamefont {Austen}\
		\bibnamefont {Lamacraft}},\ }\bibfield  {title} {\enquote {\bibinfo {title}
		{Noisy coupled qubits: Operator spreading and the fredrickson-andersen
			model},}\ }\href {http://arxiv.org/abs/1806.01723} {\bibfield  {journal}
	{\bibinfo  {journal} {{arXiv}:1806.01723}\ } (\bibinfo {year}
	{2018})}\BibitemShut {NoStop}%
\bibitem [{\citenamefont {Gopalakrishnan}\ \emph {et~al.}(2017)\citenamefont
	{Gopalakrishnan}, \citenamefont {Islam},\ and\ \citenamefont
	{Knap}}]{gopalakrishnan_noise-induced_2017}%
\BibitemOpen
\bibfield  {author} {\bibinfo {author} {\bibfnamefont {Sarang}\ \bibnamefont
		{Gopalakrishnan}}, \bibinfo {author} {\bibfnamefont {K.~Ranjibul}\
		\bibnamefont {Islam}}, \ and\ \bibinfo {author} {\bibfnamefont {Michael}\
		\bibnamefont {Knap}},\ }\bibfield  {title} {\enquote {\bibinfo {title}
		{Noise-induced subdiffusion in strongly localized quantum systems},}\ }\href
{\doibase 10.1103/PhysRevLett.119.046601} {\bibfield  {journal} {\bibinfo
		{journal} {Phys. Rev. Lett.}\ }\textbf {\bibinfo {volume} {119}},\ \bibinfo
	{pages} {046601} (\bibinfo {year} {2017})}\BibitemShut {NoStop}%
\bibitem [{\citenamefont {Amir}\ \emph {et~al.}(2009)\citenamefont {Amir},
	\citenamefont {Lahini},\ and\ \citenamefont {Perets}}]{amir_classical_2009}%
\BibitemOpen
\bibfield  {author} {\bibinfo {author} {\bibfnamefont {Ariel}\ \bibnamefont
		{Amir}}, \bibinfo {author} {\bibfnamefont {Yoav}\ \bibnamefont {Lahini}}, \
	and\ \bibinfo {author} {\bibfnamefont {Hagai~B.}\ \bibnamefont {Perets}},\
}\bibfield  {title} {\enquote {\bibinfo {title} {Classical diffusion of a
		quantum particle in a noisy environment},}\ }\href {\doibase
10.1103/PhysRevE.79.050105} {\bibfield  {journal} {\bibinfo  {journal} {Phys.
		Rev. E}\ }\textbf {\bibinfo {volume} {79}},\ \bibinfo {pages} {050105}
(\bibinfo {year} {2009})}\BibitemShut {NoStop}%
\bibitem [{\citenamefont {Breuer}\ and\ \citenamefont
	{Petruccione}(2002)}]{breuer_theory_2002}%
\BibitemOpen
\bibfield  {author} {\bibinfo {author} {\bibfnamefont {{Heinz-Peter}}\
		\bibnamefont {Breuer}}\ and\ \bibinfo {author} {\bibfnamefont {Francesco}\
		\bibnamefont {Petruccione}},\ }\href@noop {} {\emph {\bibinfo {title} {The
			Theory Of Open Quantum Systems}}}\ (\bibinfo  {publisher} {Oxford University
	Press},\ \bibinfo {address} {Oxford},\ \bibinfo {year} {2002})\BibitemShut
{NoStop}%
\bibitem [{\citenamefont {Pr{\"a}hofer}\ and\ \citenamefont
	{Spohn}(2000{\natexlab{a}})}]{prahofer_universal_2000}%
\BibitemOpen
\bibfield  {author} {\bibinfo {author} {\bibfnamefont {Michael}\ \bibnamefont
		{Pr{\"a}hofer}}\ and\ \bibinfo {author} {\bibfnamefont {Herbert}\
		\bibnamefont {Spohn}},\ }\bibfield  {title} {\enquote {\bibinfo {title}
		{Universal distributions for growth processes in 1+1 dimensions and random
			matrices},}\ }\href {\doibase 10.1103/PhysRevLett.84.4882} {\bibfield
	{journal} {\bibinfo  {journal} {Phys. Rev. Lett.}\ }\textbf {\bibinfo
		{volume} {84}},\ \bibinfo {pages} {4882--4885} (\bibinfo {year}
	{2000}{\natexlab{a}})}\BibitemShut {NoStop}%
\bibitem [{\citenamefont {Pr{\"a}hofer}\ and\ \citenamefont
	{Spohn}(2000{\natexlab{b}})}]{prahofer_statistical_2000}%
\BibitemOpen
\bibfield  {author} {\bibinfo {author} {\bibfnamefont {Michael}\ \bibnamefont
		{Pr{\"a}hofer}}\ and\ \bibinfo {author} {\bibfnamefont {Herbert}\
		\bibnamefont {Spohn}},\ }\bibfield  {title} {\enquote {\bibinfo {title}
		{Statistical self-similarity of one-dimensional growth processes},}\ }\href
{\doibase 10.1016/S0378-4371(99)00517-8} {\bibfield  {journal} {\bibinfo
		{journal} {Physica A (Amsterdam)}\ }\textbf {\bibinfo {volume} {279}},\
	\bibinfo {pages} {342--352} (\bibinfo {year}
	{2000}{\natexlab{b}})}\BibitemShut {NoStop}%
\bibitem [{\citenamefont {Cai}\ and\ \citenamefont
	{Barthel}(2013)}]{cai_algebraic_2013}%
\BibitemOpen
\bibfield  {author} {\bibinfo {author} {\bibfnamefont {Zi}~\bibnamefont
		{Cai}}\ and\ \bibinfo {author} {\bibfnamefont {Thomas}\ \bibnamefont
		{Barthel}},\ }\bibfield  {title} {\enquote {\bibinfo {title} {Algebraic
			versus exponential decoherence in dissipative many-particle systems},}\
}\href {\doibase 10.1103/PhysRevLett.111.150403} {\bibfield  {journal}
{\bibinfo  {journal} {Phys. Rev. Lett.}\ }\textbf {\bibinfo {volume} {111}},\
\bibinfo {pages} {150403} (\bibinfo {year} {2013})}\BibitemShut {NoStop}%
\bibitem [{\citenamefont {Lesanovsky}\ and\ \citenamefont
	{Garrahan}(2013)}]{lesanovsky_kinetic_2013}%
\BibitemOpen
\bibfield  {author} {\bibinfo {author} {\bibfnamefont {Igor}\ \bibnamefont
		{Lesanovsky}}\ and\ \bibinfo {author} {\bibfnamefont {Juan~P.}\ \bibnamefont
		{Garrahan}},\ }\bibfield  {title} {\enquote {\bibinfo {title} {Kinetic
			constraints, hierarchical relaxation, and onset of glassiness in strongly
			interacting and dissipative rydberg gases},}\ }\href {\doibase
	10.1103/PhysRevLett.111.215305} {\bibfield  {journal} {\bibinfo  {journal}
		{Phys. Rev. Lett.}\ }\textbf {\bibinfo {volume} {111}},\ \bibinfo {pages}
	{215305} (\bibinfo {year} {2013})}\BibitemShut {NoStop}%
\bibitem [{\citenamefont {Chaikin}\ and\ \citenamefont
	{Lubensky}(2000)}]{chaikin_principles_2000}%
\BibitemOpen
\bibfield  {author} {\bibinfo {author} {\bibfnamefont {P.~M.}\ \bibnamefont
		{Chaikin}}\ and\ \bibinfo {author} {\bibfnamefont {T.~C.}\ \bibnamefont
		{Lubensky}},\ }\href@noop {} {\emph {\bibinfo {title} {Principles of
			Condensed Matter Physics}}}\ (\bibinfo  {publisher} {Cambridge University
	Press},\ \bibinfo {address} {Cambridge; New York, {USA}},\ \bibinfo {year}
{2000})\BibitemShut {NoStop}%
\bibitem [{\citenamefont {Bauer}\ \emph {et~al.}(2017)\citenamefont {Bauer},
	\citenamefont {Bernard},\ and\ \citenamefont {Jin}}]{bauer_stochastic_2017}%
\BibitemOpen
\bibfield  {author} {\bibinfo {author} {\bibfnamefont {Michel}\ \bibnamefont
		{Bauer}}, \bibinfo {author} {\bibfnamefont {Denis}\ \bibnamefont {Bernard}},
	\ and\ \bibinfo {author} {\bibfnamefont {Tony}\ \bibnamefont {Jin}},\
}\bibfield  {title} {\enquote {\bibinfo {title} {Stochastic dissipative
		quantum spin chains (i) : Quantum fluctuating discrete hydrodynamics},}\
}\href {\doibase 10.21468/SciPostPhys.3.5.033} {\bibfield  {journal}
{\bibinfo  {journal} {{SciPost} Physics}\ }\textbf {\bibinfo {volume} {3}},\
\bibinfo {pages} {033} (\bibinfo {year} {2017})}\BibitemShut {NoStop}%
\bibitem [{\citenamefont {Han}\ and\ \citenamefont
	{Hartnoll}(2018)}]{han_locality_2018}%
\BibitemOpen
\bibfield  {author} {\bibinfo {author} {\bibfnamefont {Xizhi}\ \bibnamefont
		{Han}}\ and\ \bibinfo {author} {\bibfnamefont {Sean~A.}\ \bibnamefont
		{Hartnoll}},\ }\bibfield  {title} {\enquote {\bibinfo {title} {Locality bound
			for dissipative quantum transport},}\ }\href {\doibase
	10.1103/PhysRevLett.121.170601} {\bibfield  {journal} {\bibinfo  {journal}
		{Phys. Rev. Lett.}\ }\textbf {\bibinfo {volume} {121}},\ \bibinfo {pages}
	{170601} (\bibinfo {year} {2018})}\BibitemShut {NoStop}%
\bibitem [{\citenamefont {\ifmmode \check{Z}\else
		\v{Z}\fi{}nidari\ifmmode~\check{c}\else
		\v{c}\fi{}}(2010)}]{znidaric_dephasing-induced_2010}%
\BibitemOpen
\bibfield  {author} {\bibinfo {author} {\bibfnamefont {Marko}\ \bibnamefont
		{\ifmmode \check{Z}\else \v{Z}\fi{}nidari\ifmmode~\check{c}\else
			\v{c}\fi{}}},\ }\bibfield  {title} {\enquote {\bibinfo {title}
		{Dephasing-induced diffusive transport in the anisotropic heisenberg
			model},}\ }\href {\doibase 10.1088/1367-2630/12/4/043001} {\bibfield
	{journal} {\bibinfo  {journal} {New J. Phys.}\ }\textbf {\bibinfo {volume}
		{12}},\ \bibinfo {pages} {043001} (\bibinfo {year} {2010})}\BibitemShut
{NoStop}%
\bibitem [{\citenamefont {Lieb}\ and\ \citenamefont
	{Robinson}(1972)}]{lieb_finite_1972}%
\BibitemOpen
\bibfield  {author} {\bibinfo {author} {\bibfnamefont {Elliott~H.}\
		\bibnamefont {Lieb}}\ and\ \bibinfo {author} {\bibfnamefont {Derek~W.}\
		\bibnamefont {Robinson}},\ }\bibfield  {title} {\enquote {\bibinfo {title}
		{The finite group velocity of quantum spin systems},}\ }\href {\doibase
	10.1007/BF01645779} {\bibfield  {journal} {\bibinfo  {journal} {Commun.Math.
			Phys.}\ }\textbf {\bibinfo {volume} {28}},\ \bibinfo {pages} {251--257}
	(\bibinfo {year} {1972})}\BibitemShut {NoStop}%
\bibitem [{\citenamefont {Deissler}(1984)}]{deissler_one-dimensional_1984}%
\BibitemOpen
\bibfield  {author} {\bibinfo {author} {\bibfnamefont {Robert~J.}\
		\bibnamefont {Deissler}},\ }\bibfield  {title} {\enquote {\bibinfo {title}
		{One-dimensional strings, random fluctuations, and complex chaotic
			structures},}\ }\href {\doibase 10.1016/0375-9601(84)90823-5} {\bibfield
	{journal} {\bibinfo  {journal} {Physics Letters A}\ }\textbf {\bibinfo
		{volume} {100}},\ \bibinfo {pages} {451--454} (\bibinfo {year}
	{1984})}\BibitemShut {NoStop}%
\bibitem [{\citenamefont {Kaneko}(1986)}]{kaneko_lyapunov_1986}%
\BibitemOpen
\bibfield  {author} {\bibinfo {author} {\bibfnamefont {Kunihiko}\
		\bibnamefont {Kaneko}},\ }\bibfield  {title} {\enquote {\bibinfo {title}
		{Lyapunov analysis and information flow in coupled map lattices},}\ }\href
{\doibase 10.1016/0167-2789(86)90149-1} {\bibfield  {journal} {\bibinfo
		{journal} {Physica D: Nonlinear Phenomena}\ }\textbf {\bibinfo {volume}
		{23}},\ \bibinfo {pages} {436--447} (\bibinfo {year} {1986})}\BibitemShut
{NoStop}%
\bibitem [{\citenamefont {Peschel}(2003)}]{peschel_calculation_2003}%
\BibitemOpen
\bibfield  {author} {\bibinfo {author} {\bibfnamefont {Ingo}\ \bibnamefont
		{Peschel}},\ }\bibfield  {title} {\enquote {\bibinfo {title} {Calculation of
			reduced density matrices from correlation functions},}\ }\href {\doibase
	10.1088/0305-4470/36/14/101} {\bibfield  {journal} {\bibinfo  {journal} {J.
			Phys. A: Math. Gen.}\ }\textbf {\bibinfo {volume} {36}},\ \bibinfo {pages}
	{L205} (\bibinfo {year} {2003})}\BibitemShut {NoStop}%
\end{thebibliography}

%

\end{document}